\documentclass[aps,pre,showpacs,amsmath,amssymb,amsfonts,lengthcheck,twocolumn,longbibliography,superscriptaddress]{revtex4-2}

\usepackage{changes} 

\usepackage{graphicx}
\usepackage{subfigure}
\usepackage{amsthm}
\usepackage{amsmath}
\usepackage{verbatim}
\usepackage{dcolumn}
\usepackage{bm}
\usepackage{color}
\usepackage[colorlinks=true,citecolor=blue,linkcolor=blue,urlcolor=blue]{hyperref}%
\usepackage{xcolor}
\usepackage{dsfont}
\usepackage{longtable}
\allowdisplaybreaks

\newcommand{\bla}{bla\\bla\\bla\\bla\\bla}

\begin{document}

\title{Multilayer Graphene as an Endoreversible Otto Engine}

\author{Nathan M Myers}
\email{myersn1@vt.edu}
\affiliation{Department of Physics, Virginia Tech, Blacksburg, Virginia 24061, USA}
\author{Francisco J. Pe\~na}
\email{francisco.penar@usm.cl}
\affiliation{Departamento de F\'isica, Universidad T\'ecnica Federico Santa Mar\'ia, Av. Espa\~na 1680, Valpara\'iso 11520, Chile}
\affiliation{Millennium Nucleus in NanoBioPhysics (NNBP), Av. Espa\~na 1680, Valpara\'iso 11520, Chile;}
\author{Natalia Cortés}
\affiliation{Instituto de Alta Investigación, Universidad de Tarapacá, Casilla 7D, Arica, Chile}
\affiliation{Department of Physics and Astronomy, and Nanoscale and Quantum Phenomena Institute, Ohio University, Athens, Ohio 45701, USA}
\author{Patricio Vargas}
\affiliation{Departamento de F\'isica, CEDENNA, Universidad T\'ecnica Federico Santa Mar\'ia,  Av. Espa\~na 1680, Valpara\'iso 11520, Chile}

\date{\today}

\begin{abstract}
Graphene is perhaps the most prominent ``Dirac material,” a class of systems whose electronic structure gives rise to charge carriers that behave as relativistic fermions. In multilayer graphene several crystal sheets are stacked such that the honeycomb lattice of each layer is displaced along one of the lattice edges. When subject to an external magnetic field, the scaling of the multilayer energy spectrum with the magnetic field, and thus the system's thermodynamic behavior, depends strongly on the number of layers. With this in mind, we examine the performance of a finite-time endoreversible Otto cycle with multilayer graphene as its working medium. We show that there exists a simple relationship between the engine efficiency and the number of layers, and that the efficiency at maximum power can exceed that of a classical endoreversible Otto cycle.       
\end{abstract}

\maketitle

\section{\label{sec:1} Introduction}

There is rapidly growing interest in the development of quantum technologies, devices that take advantage of the unique properties of quantum systems to enhance their performance. This, in turn, has led to  increased focus on the field of quantum thermodynamics \cite{deffner2019quantum, gemmer2009quantum, kosloff2013quantum, vinjanampathy2016quantum}. Within the broad spectrum of topics that fall under the umbrella of quantum thermodynamics, significant attention is paid to the study of quantum heat engines - devices that extend the principles of classical heat engines to include working mediums made up of quantum systems \cite{Myers2022, Kosloff2014, quan2007quantum, quan2009quantum, palao2001quantum}.

Efficiency, defined as the ratio of the net work to the heat absorbed from the hot reservoir, is by far the most prominent metric of engine performance. To maximize engine efficiency the strokes of the cycle must be carried out quasistatically. However, truly quasistatic strokes require infinite time to implement, thus leading to vanishing power output. Practically useful metrics of heat engine performance must therefore account for cycles implemented in finite time. \textit{Endoreversible thermodynamics} \cite{Curzon1975, Rubin1979, Hoffmann1997} provides a framework for introducing finite-time behavior by assuming that, while the working medium remains in a state of local equilibrium at all times during the cycle, the heating and cooling strokes occur quickly enough that the working medium never fully thermalizes with the hot and cold reservoirs. A prominent performance characteristic within endoreversible thermodynamics is the \textit{efficiency at maximum power} (EMP) which corresponds to maximizing the power output with respect to the external control parameter and then determining the efficiency at that maximum power output. Endoreversible cycles have also been studied in the context of quantum heat engines. We draw particular attention to Ref. \cite{deffner2018efficiency}, where it was shown that the EMP of an endoreversible Otto cycle with a quantum harmonic oscillator as the working medium exceeds the Curzorn-Albhorn (CA) efficiency, the EMP achieved by the Otto cycle with a classical working medium. 

When considering possible systems to serve as the working medium of a quantum heat engine, graphene stands out as intriguing candidate. Graphene's optical, electronic, and mechanical properties have been extensively studied in recent years \cite{yin2017landau, yang2018structure, choi2010synthesis, si2016strain, novoselov2012roadmap, ho2010electronic, Allen2010}. Furthermore, graphene is a prominent Dirac material, systems whose low energy excitations behave as relativistic massless  fermions \cite{Wehling2014}. Thus the study of quantum heat engines with graphene as a working medium can give insight into the role of relativistic quantum features in engine performance \cite{Munoz2012, Myers2021NJP}. In particular, the performance of a quasistatic Otto engine with twisted bilayer graphene was recently studied \cite{singh2021magic}. In this work it was found that the highest efficiency is reached when the twist angle corresponds to the magic angle of 0.96 degrees. These results show that for a heat engine with multilayer graphene as the working medium the number and configuration of the crystal sheets plays a significant role in the engine performance. Significant attention has also been given to graphene-based engines in the context of continuous, thermoelectric machines \cite{Karbaschi2019, Mani2017, Mani2019, Mani2019PCC}. Notably, graphene has also been used in the construction of an experimental nanoscale cyclic heat engine \cite{Lee2014}.

In this manuscript we analyze the finite time performance of an Otto cycle with multilayer graphene as the working medium using the framework of endoreversible thermodynamics. In section \ref{sec:2} we provide relevant background, including the analytical results of the energy spectrum for monolayer, bilayer, and trilayer graphene. In section \ref{sec:3} we determine a closed form for the partition function and examine the equilibrium thermodynamic behavior of multilayer graphene. In section \ref{sec:4} we introduce the endoreversible Otto cycle for multilayer graphene before presenting the results for the engine efficiency, power output, and EMP in section \ref{sec:5}.  

\section{Model}
\label{sec:2}

In multilayer graphene, the crystal sheets are placed on top of each other in different stacking configurations and are connected through weak van der Waals forces. The stacking configurations are determined by the orientation of the two triangular sublattices that make up the primary honeycomb lattice of a single sheet. For two stacked sheets, three possible orientations, A, B, and C, are possible, each corresponding to displacing one of the sublattice atoms along the edge of the honeycomb with respect to the neighboring sheet \cite{min2008electronic}. Subject to a perpendicular external magnetic field, these systems can be analyzed using a $\pi$-orbital continuum model. Such an analysis is described extensively in Ref. \cite{min2008electronic}. In our analysis we will focus on two particular stacking configurations. For bilayer graphene we consider Bernal stacking, also known as AB stacking. For the case of trilayer graphene we consider the rhombohedral configuration, also known as ABC stacking \cite{geisenhof2019anisotropic}. Significantly, an exact analytical result for the energy spectrum as a function of the external magnetic field can be found for these two cases.

\subsection{Monolayer Graphene}

In monolayer graphene the application of a perpendicular magnetic field results in unevenly spaced Landau levels with an energy spectrum proportional to the root of the level quantum number $n$ \cite{Miller2009}, 
\begin{equation}
\label{singlelayerenergy}
    E_{n}= \pm \sqrt{2e\hbar v_{f}^{2} n B}, \, n=0, 1, 2, ..., 
\end{equation}
where $B$ is the magnitude of the magnetic field, $e$ is the electron charge, $\hbar$ is Planck's constant, and $v_{f}$ is the Fermi velocity ($\sim 10^{6}$ m/s). Such an energy spectrum is characteristic of ultra-relativistic massless particles with the Fermi velocity playing the role of the speed of light. The positive energy branch corresponds to particle behavior and the negative energy branch to holes \cite{min2008electronic}. These energy levels are four times degenerate, including the zero energy state, where the factor of four arises from spin degeneracy and non-equivalent BZ points $K$ and $K^{'}$, known as valley degeneracy.

\subsection{Bilayer Graphene: AB stacking}

For a bilayer system, the most stable coupling corresponds to Bernal, or AB, stacking. This consists of displacing the A sublattice atoms of the upper layer so that they lie on top of the B sublattice atoms of the lower layer. Notably, the bilayer system has a quadratic dispersion relation, which gives rise to an interesting phenomenon. While the Dirac equation still models the dynamics of the low energy states, the quadratic dispersion relation indicates that the described charge carriers have mass. In this case, under a perpendicular external magnetic field, the low energy spectrum takes the form \cite{McCann2006},
\begin{equation}
\label{Bernal}
    E_{n}= \pm \hbar \omega_{c} \sqrt{n(n-1)}, \; n= 0, 1, 2,..
\end{equation}
where $\omega_{c} \equiv e B/m^{*}$ corresponds to the cyclotron frequency.  This effective mass, $m^{*}$, is related to the Fermi velocity and the interlayer interaction parameter, $t_{\perp}$, by $m^{*}=t_{\perp}/2 v_{f}^{2}$. This corresponds to a numerical value of $m^{*} \sim (0.039 \pm 0.002)\; m_{e}$, where $m_{e}$ is the electron rest mass. Note that Eq.~(\ref{Bernal}) has two zero energy levels, corresponding to $n=0$ and $n=1$.

\subsection{ Trilayer Graphene: ABC stacking}

Trilayer graphene in the ABC configuration acts as a semiconductor with a gate tunable band gap. The energy spectrum has the form \cite{min2008electronic},
\begin{equation}
\label{rhombo}
    E_{n}= \pm \frac{\left(2\hbar v_{F}^{2} e B\right)^{3/2}}{t_{\perp}^{2}}\sqrt{n(n-1)(n-2)}, \; n=0, 1, 2, ...
\end{equation}
Note that in the case of trilayer graphene the zero-energy state is 12 fold degenerate, while the other energy states remain fourfold degenerate just as in the bilayer and monolayer case. 

\begin{figure}
	\subfigure[]{
		\includegraphics[width=.43\textwidth]{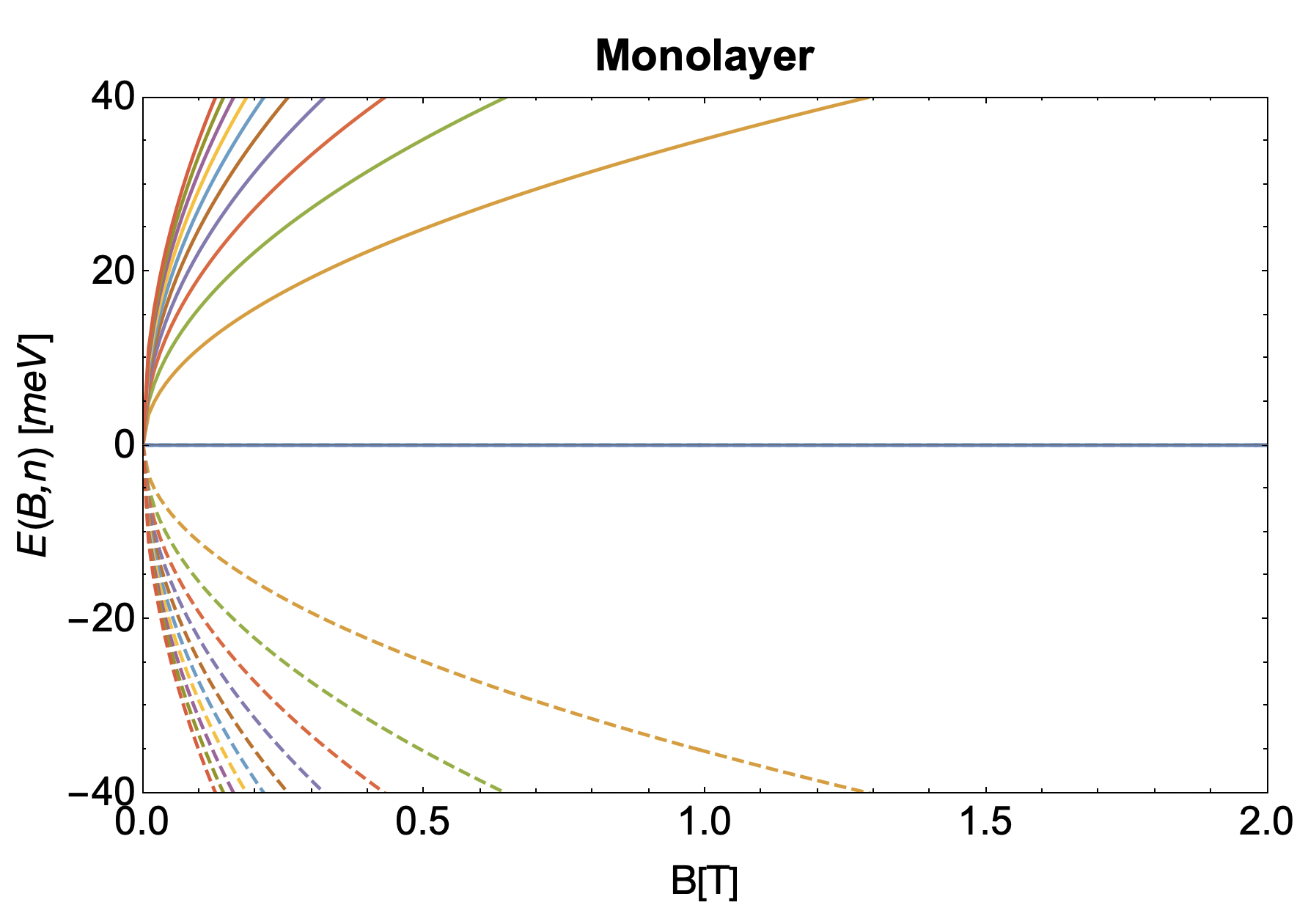}
	}
	\hspace{5mm}
	\subfigure[]{
		\includegraphics[width=.43\textwidth]{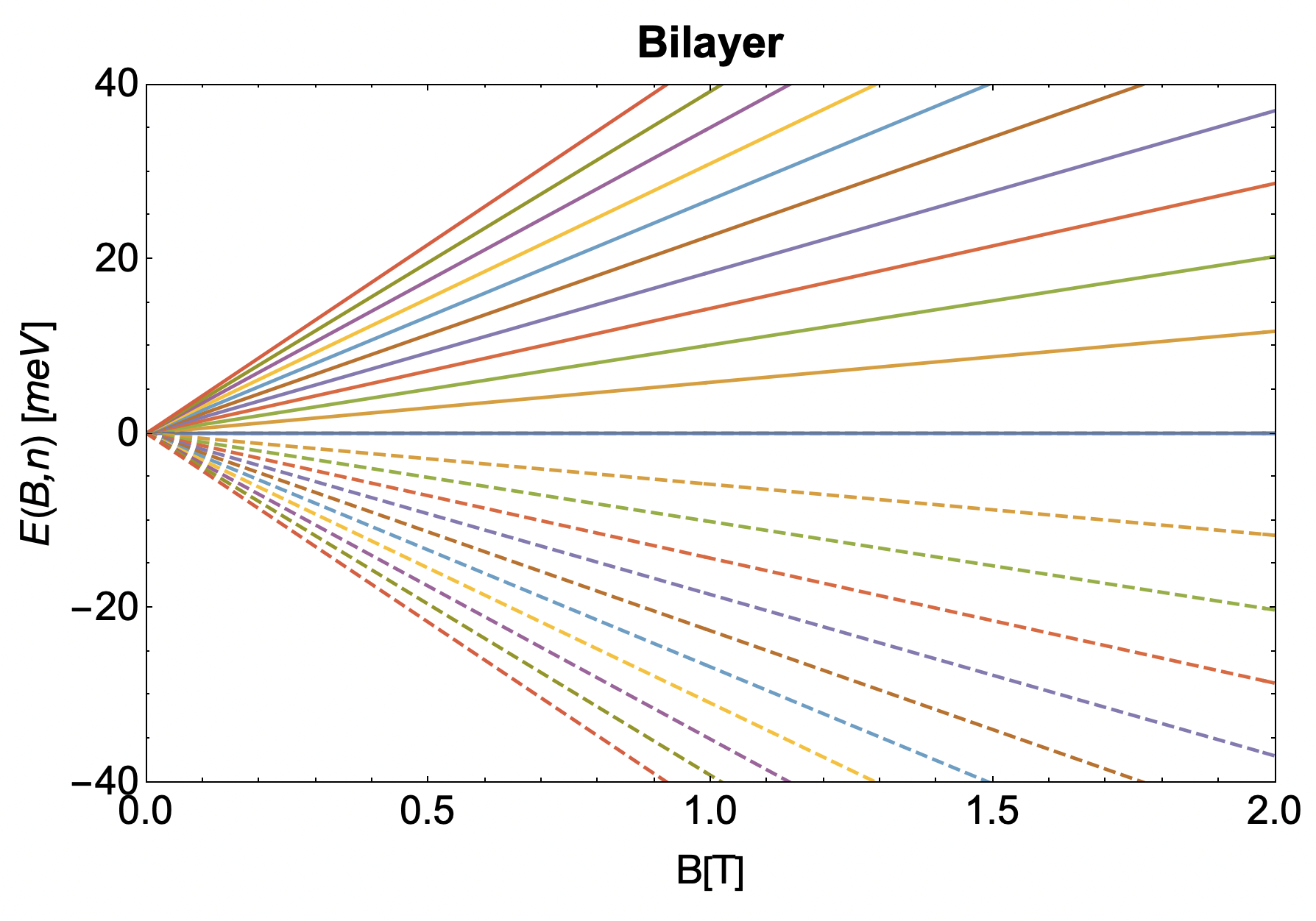}
	}
	\hspace{5mm}
	\subfigure[]{
		\includegraphics[width=.43\textwidth]{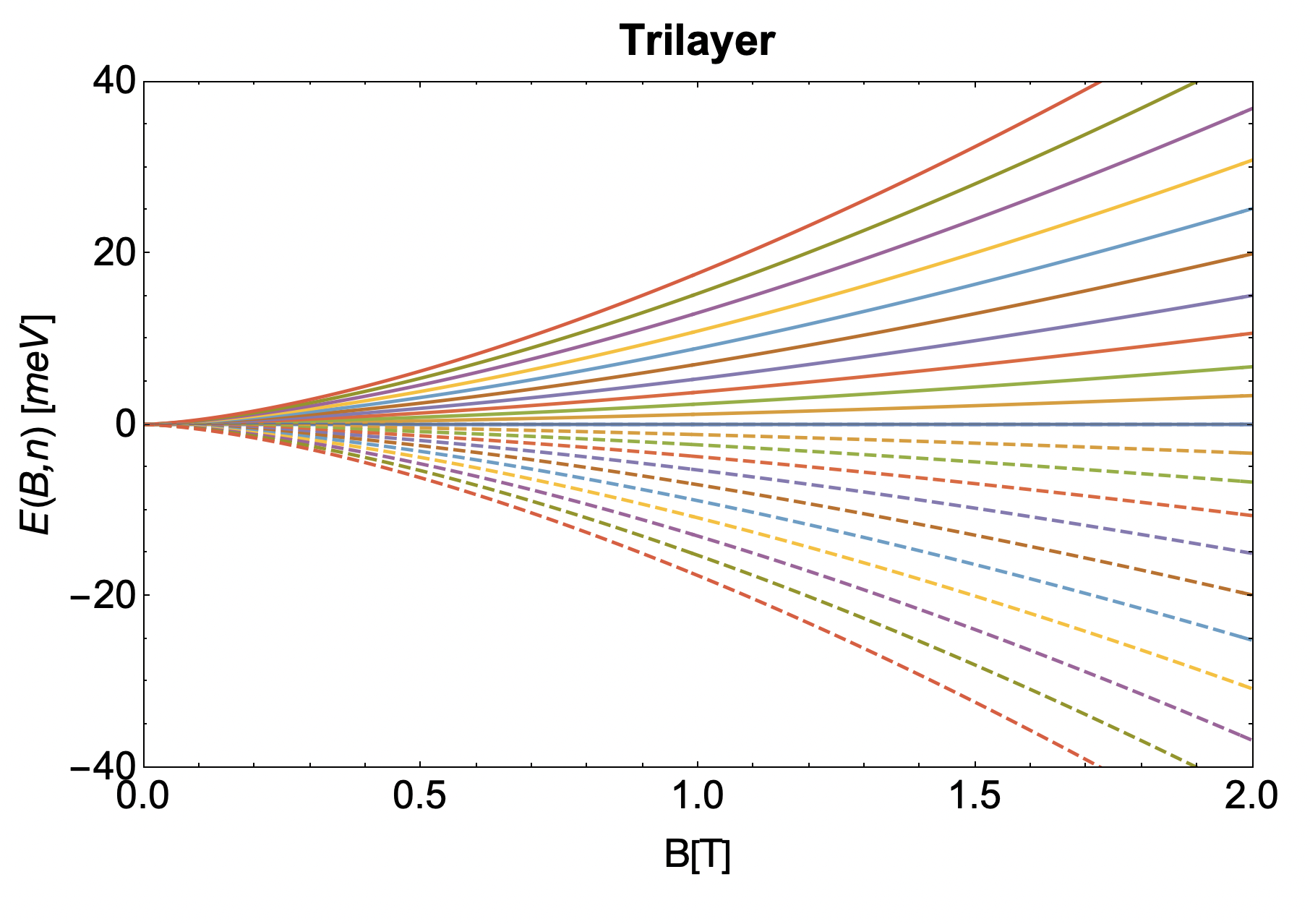}
	}
    \caption{Energy spectrum as a function of the external magnetic field for the first 10 Landau levels of (a) monolayer, (b) bilayer (with $m^{*}=0.03 \; m_{e}$) and (c) trilayer graphene. Solid (dashed) lines correspond to electrons (holes).}
    \label{spectrummono-bi-tri}
\end{figure}


\section{\label{sec:3} Partition Function and Equilibrium Thermodynamics}

Comparing the energy spectra presented in the previous section, we see a pattern emerging in how the energy scales with the magnetic field. Each energy is proportional to $B^{\mathcal{N}/2}$ where $\mathcal{N}$ is the number of layers. To illustrate this behavior, we plot the first ten positive and negative energy states as a function of the external field for monolayer, bilayer, and trilayer graphene in Fig. \ref{spectrummono-bi-tri}. We also see the energy spectra follow a common structure in regard to the quantum number $n$, which takes the form,
\begin{equation}
    f_{\mathcal{N}}(n)=\sqrt{\displaystyle\prod_{k=0}^{\mathcal{N}-1} \left( n - k \right)}.
\end{equation}
Therefore, we can compactly write the energy spectrum for the multilayer system in the form \cite{min2008electronic},
\begin{equation}
\label{simplyenergy}
    \mathcal{E}_{n,\mathcal{N}}=\theta_{\mathcal{N}}B^{\frac{\mathcal{N}}{2}}f_{\mathcal{N}}(n),
\end{equation}
where $\theta_{\mathcal{N}} \equiv (2 e \hbar v_f^2)^{\mathcal{N}/2} (t_{\perp})^{1-\mathcal{N}}$, is a constant that depends on the number of layers and the stacking structure of the system. 

Note that the energy spectra in  Eqs. (\ref{singlelayerenergy}), (\ref{Bernal}), and (\ref{rhombo}) include both positive and negative energy solutions corresponding to particles and holes, respectively. In order to calculate the partition function for each system, we include only the positive energy branches. The selection of positive energies can be experimentally achieved by means of transport measurements focused on conduction electrons, as demonstrated in \cite{perez2020entropy}. Furthermore, to accurately determine the partition function we need to consider the degeneracy of the energy levels, especially for the zero-energy state. The compact form of counting these degenerate states in the partition function is given by,
\begin{equation}
    \label{eq:PartSum}
    \mathcal{Z} = 4 \, (\mathcal{N} - 1) + \sum_{n =0}^{\infty} 4 \, e^{-\beta \mathcal{E}_{n,\mathcal{N}}} .
\end{equation}
The partition function for the energy spectrum given in Eq.~(\ref{simplyenergy}) does not have a simple closed-form solution, except for the case of one layer. However, if we assume that the number of layers $\mathcal{N}$ is not very large compared with the number of states $n$, then the energy spectrum can be approximated as,
\begin{equation}
    \mathcal{E}_{n, \mathcal{N}} \approx \theta_{\mathcal{N}} B^{\frac{\mathcal{N}}{2}}\, n^{\frac{\mathcal{N}}{2}}.
\end{equation}

For large $n$ we can approximate the partition function sum as an integral of the form, 
\begin{equation}
    \mathcal{Z} \approx 4(\mathcal{N}-1) + 4 \int_{0}^{\infty} dn \, e^{-\beta \theta_{\mathcal{N}} B^{\frac{\mathcal{N}}{2}} \, n^{\frac{\mathcal{N}}{2}}}.
\end{equation} 
Noting that, 
\begin{equation}
    \int_{0}^{\infty} dx \, e^{-a x^{\frac{\mathcal{N}}{2}}} = a^{-\frac{2}{\mathcal{N}}} \,  \Gamma \left(\frac{2+ \mathcal{N}}{\mathcal{N}}\right),
\end{equation}
we obtain a simple analytical form for the partition function,
\begin{equation}
\label{eq:PartFcn}
\mathcal{Z}\left(T, B, \mathcal{N}\right) = 4 (\mathcal{N} - 1) + 4 \left(\frac{\theta_{\mathcal{N}} B^{\frac{\mathcal{N}}{2}}}{k_B T}\right)^{-\frac{2}{\mathcal{N}}} \,  \Gamma \left(\frac{2+ \mathcal{N}}{\mathcal{N}}\right).
\end{equation}

\begin{figure*}
	\subfigure[]{
		\includegraphics[width=.29\textwidth]{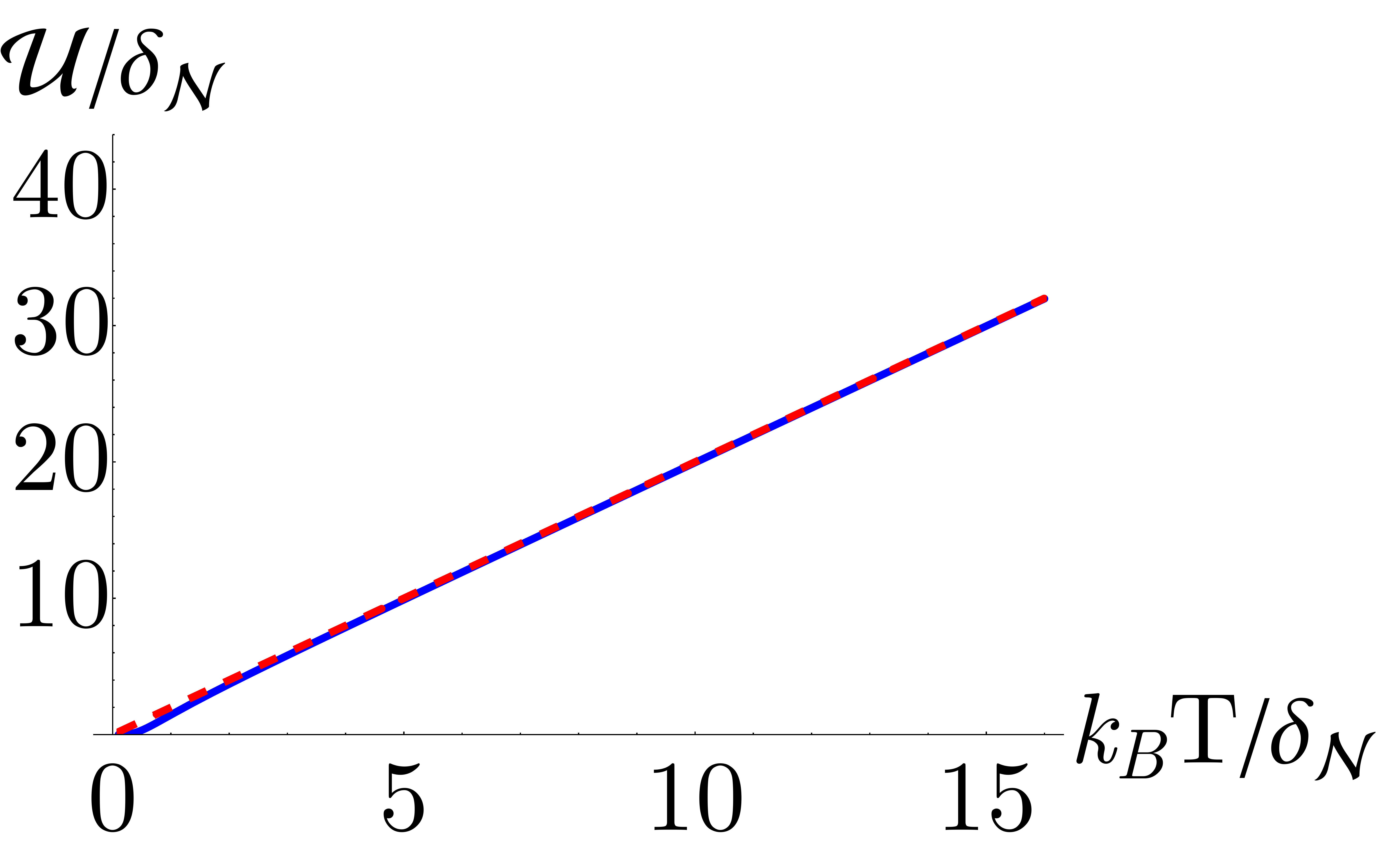}
	}
	\hspace{5mm}
	\subfigure[]{
		\includegraphics[width=.29\textwidth]{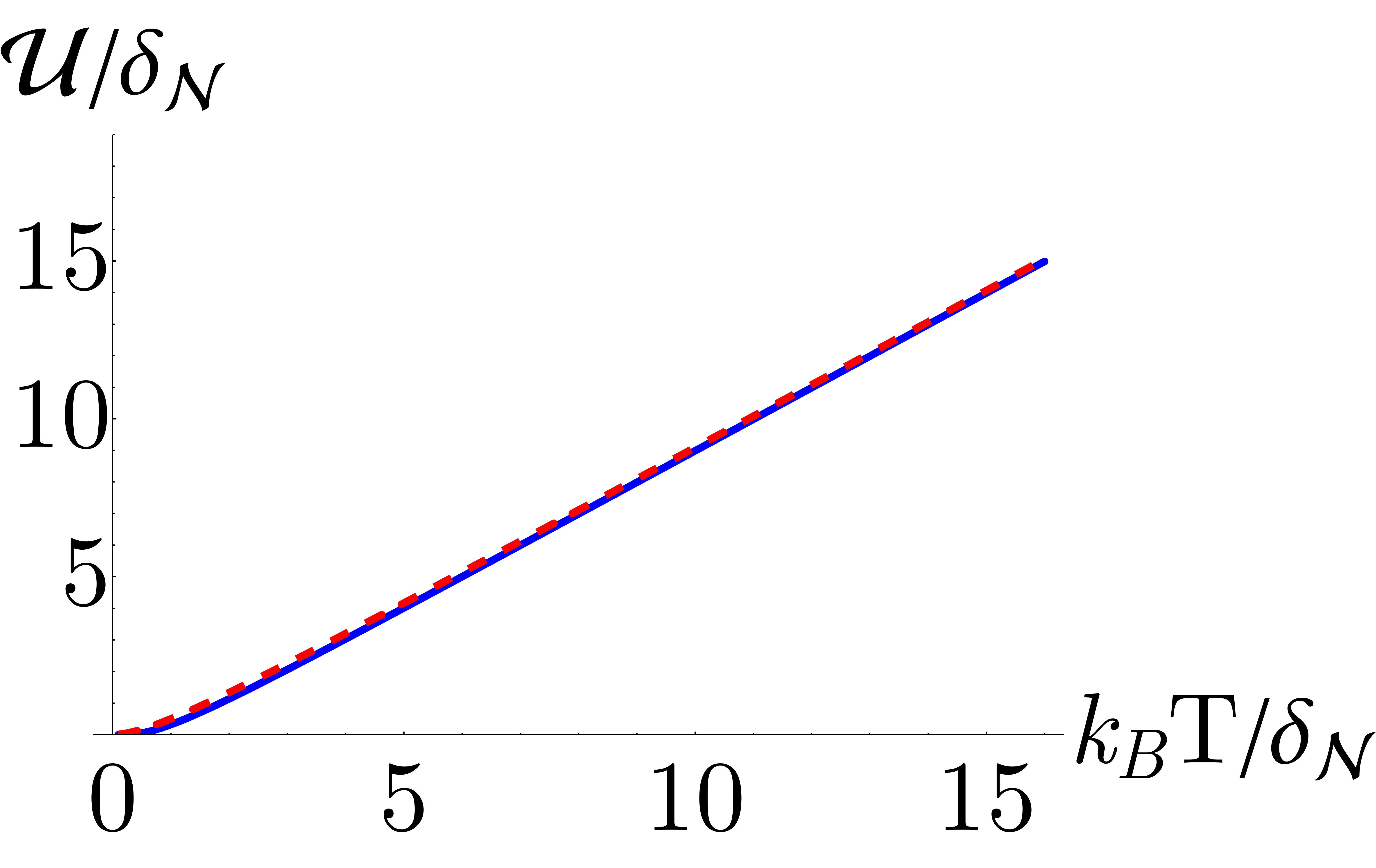}
	}
	\hspace{5mm}
	\subfigure[]{
		\includegraphics[width=.29\textwidth]{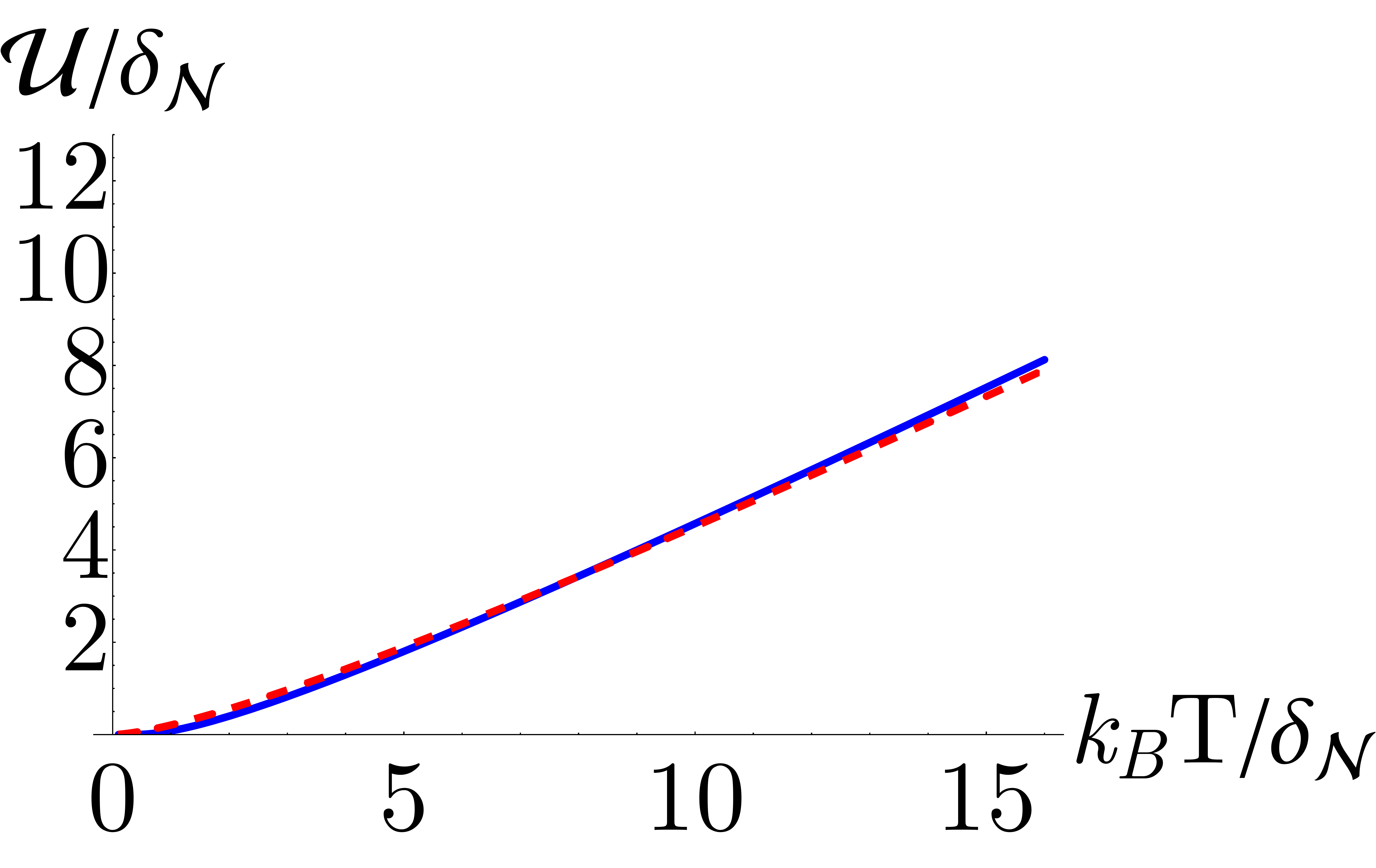}
	}
    \caption{Internal energy as a function of temperature determined from the analytical approximation for the partition function given in Eq. \eqref{eq:PartFcn} (red, dashed) and from a numerical summation obtained by truncating Eq. \eqref{eq:PartSum} after the first 50,000 terms (blue, solid) for (a) monolayer, (b) bilayer and (c) trilayer graphene. Here $\delta_{\mathcal{N}} \equiv \theta_{\mathcal{N}} B^{\mathcal{N}/2}$ such that the plot axes are unitless.}
    \label{fig:internalenergycomp}
\end{figure*}

\begin{figure*}
	\subfigure[]{
		\includegraphics[width=.29\textwidth]{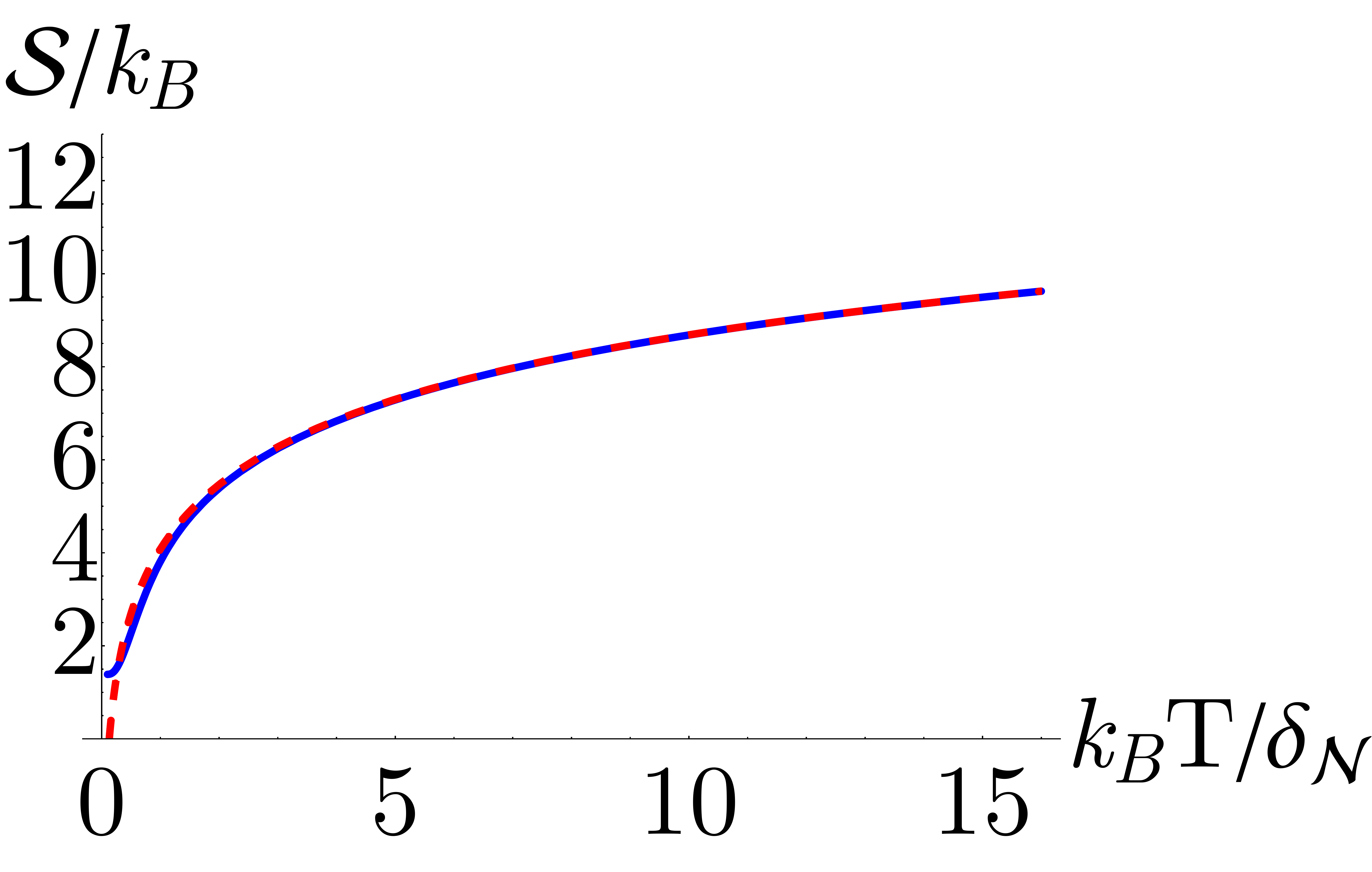}
	}
	\hspace{5mm}
	\subfigure[]{
		\includegraphics[width=.29\textwidth]{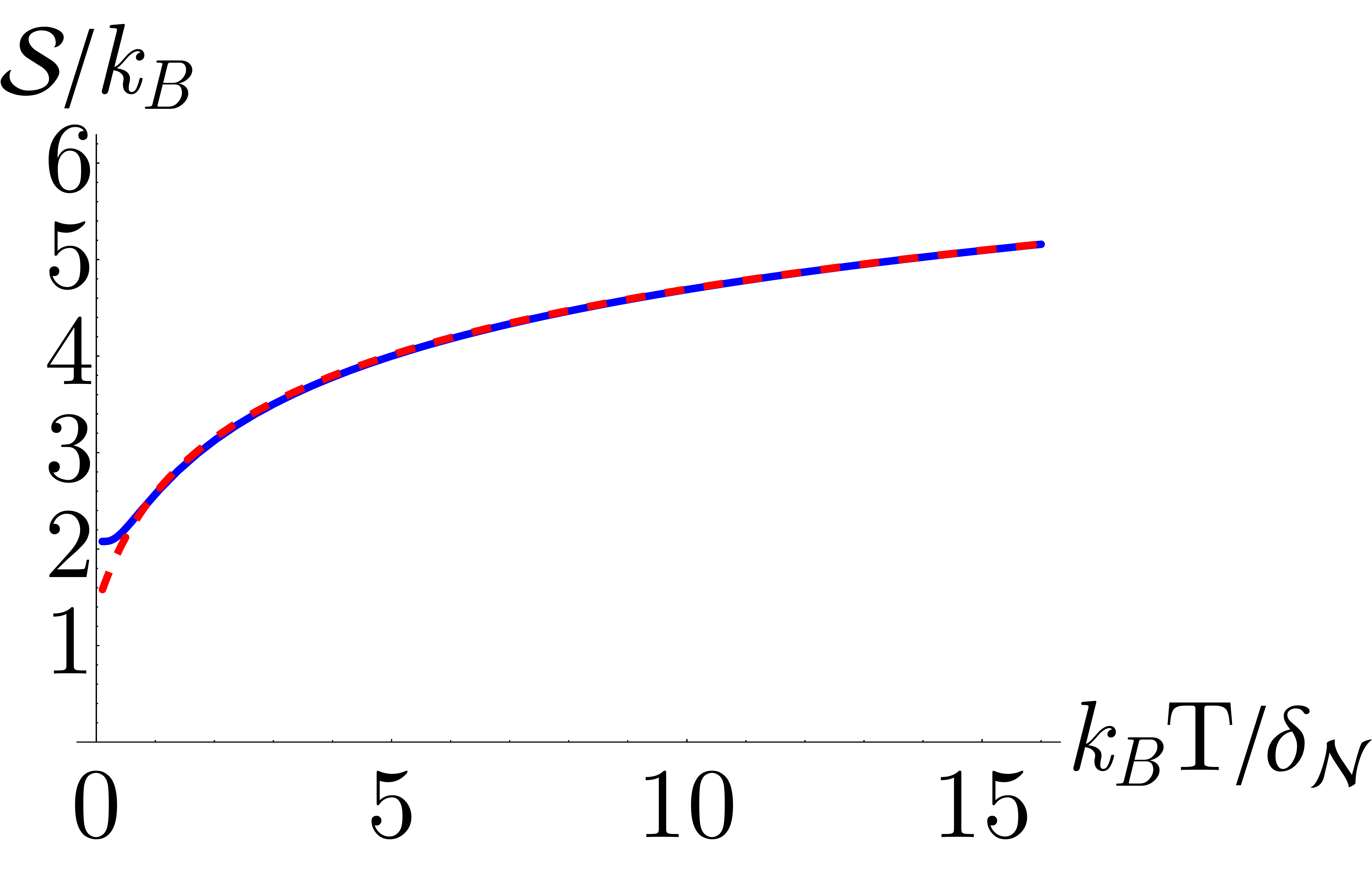}
	}
	\hspace{5mm}
	\subfigure[]{
		\includegraphics[width=.29\textwidth]{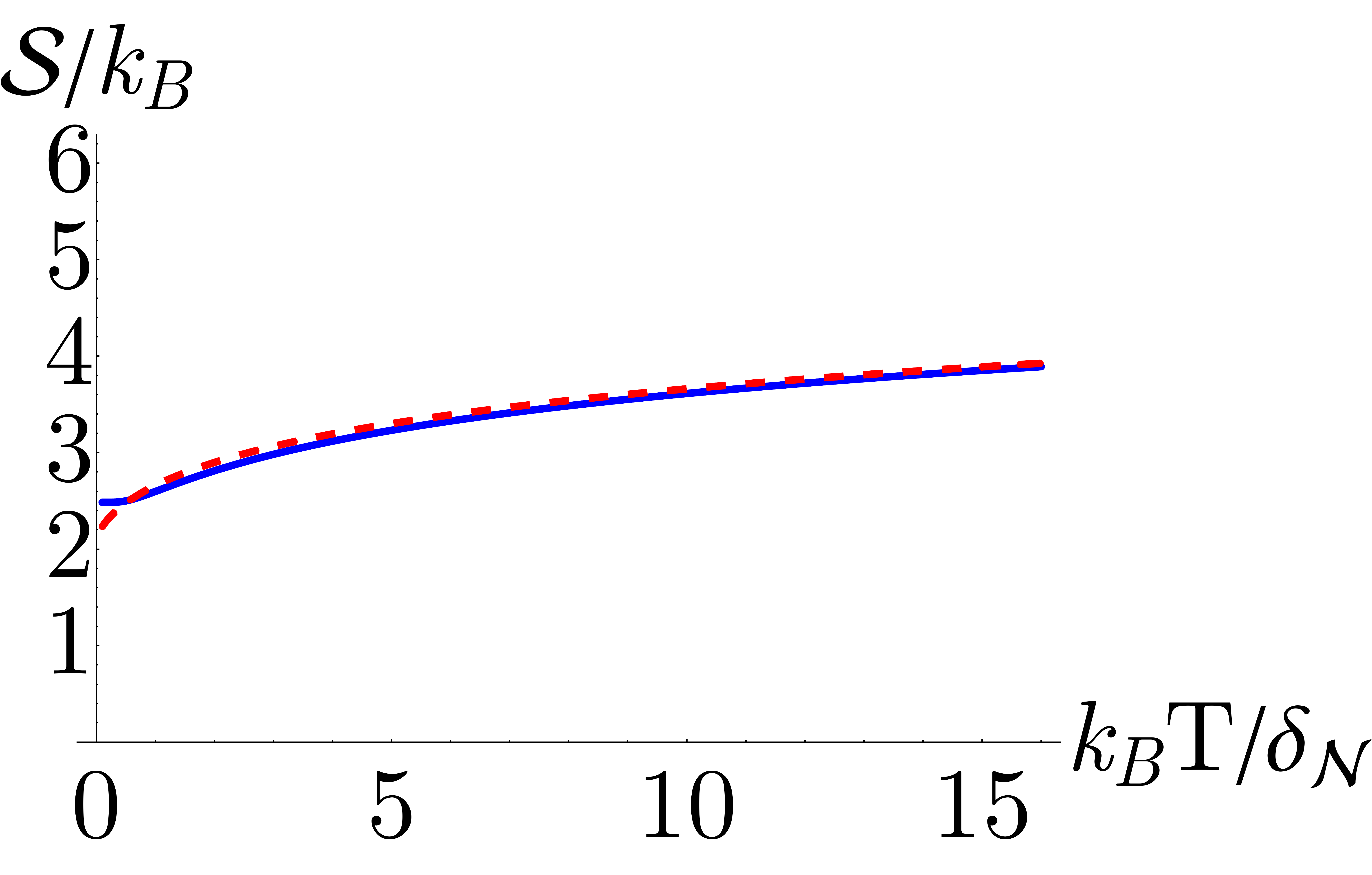}
	}
    \caption{Entropy as a function of temperature determined from the analytical approximation for the partition function given in Eq. \eqref{eq:PartFcn} (red, dashed) and from a numerical summation obtained by truncating Eq. \eqref{eq:PartSum} after the first 50,000 terms (blue, solid) for (a) monolayer, (b) bilayer and (c) trilayer graphene. Here $\delta_{\mathcal{N}} \equiv \theta_{\mathcal{N}} B^{\mathcal{N}/2}$ such that the plot axes are unitless.}
    \label{fig:entropycomp}
\end{figure*}

From the partition function, all relevant equilibrium thermodynamic properties can be determined as follows,
\begin{equation}
    \label{eq:FandS}
    \mathcal{F}= -k_{B} T \ln \mathcal{Z}, \,\, \mathcal{S} = -\left(\frac{\partial \mathcal{F}}{\partial T}\right)_{B}, 
\end{equation}
\begin{equation}
    \mathcal{U} = k_{B} T^{2} \left(\frac{\partial \ln \mathcal{Z}}{\partial T}\right)_{B}, \,\, \mathcal{C}_{B} = \left(\frac{\partial \mathcal{U}}{\partial T}\right)_{B},
\end{equation}
\begin{equation}
    \mathcal{M} = -\left(\frac{\partial \mathcal{F}}{\partial B}\right),
\end{equation}
where $\mathcal{F}$ is the free energy, $\mathcal{S}$ is the entropy, $\mathcal{U}$ is the internal energy, $\mathcal{C}_B$ is the heat capacity, and $\mathcal{M}$ is the magnetization. In Figs. \ref{fig:internalenergycomp} and \ref{fig:entropycomp} we compare the internal energy ($\mathcal{U}$) and entropy ($\mathcal{S}$), respectively, for monolayer, bilayer, and trilayer graphene. To ensure that our analytical approximation for the partition function is valid, we also plot the internal energy and entropy determined from numerical calculations of the partition function sum up to 50,000 terms.  


\section{The endoreversible Otto cycle}
\label{sec:4}

\begin{figure}
    \includegraphics[width=0.44\textwidth]{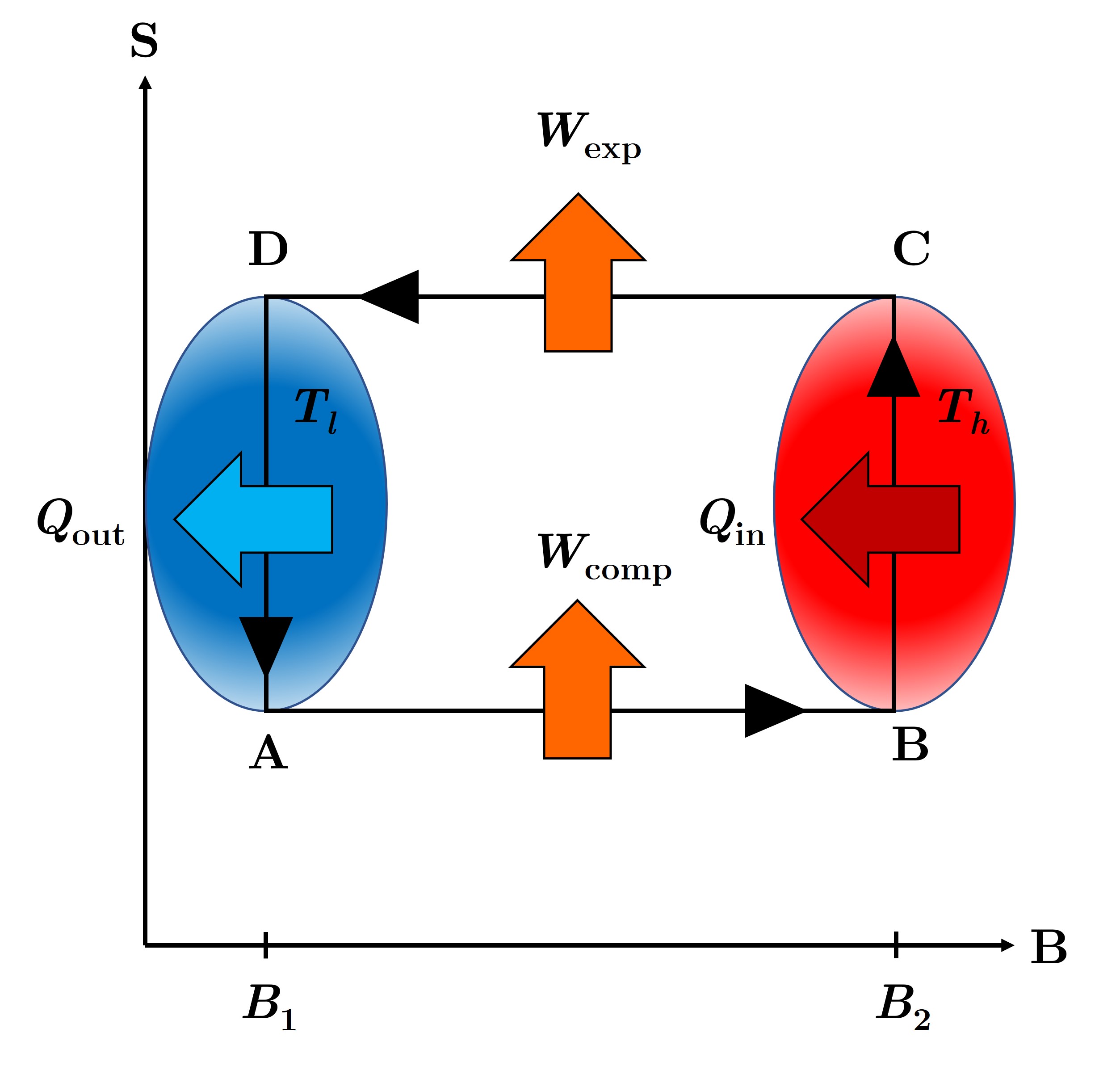}
    \caption{Entropy (S) versus external field (B) diagram for the Otto Cycle. Note that the system is only in contact with the thermal reservoirs during the isochoric (vertical) strokes. Note that in the endoreversible framework the working medium does not fully thermalize to the temperatures $T_{h}$ and $T_{l}$ of the hot and cold reservoirs  at points $\mathrm{C}$ and $\mathrm{A}$, respectively. Here $Q_{\mathrm{in}}$ is the amount of heat drawn from the hot reservoir during the heating stroke (B $\rightarrow$ C) and $Q_{\mathrm{out}}$ is the amount of heat expelled to the cold reservoir during the cooling stroke (D $\rightarrow$ A). Similarly, $W_{\mathrm{comp}}$ is the amount of work supplied to the working medium during the compression stroke (A $\rightarrow$ B), while $W_{\mathrm{exp}}$ is the amount of work extracted from the working medium during the expansion stroke (C $\rightarrow$ D).}
    \label{OttoCycle}
\end{figure}

The Otto cycle consists of four strokes, illustrated graphically in Fig. \ref{OttoCycle} using an entropy ($S$) - magnetic field ($B$) diagram. The first stroke ($\mathrm{A}\rightarrow \mathrm{B}$) is an isentropic compression in which the external field is varied from $B_1$ to $B_2$ while the working medium is isolated from the thermal reservoirs. During this stroke an amount of work, $W_{\mathrm{comp}}$, must be supplied to compress the working medium. The second stroke ($\mathrm{B}\rightarrow \mathrm{C}$) is an isochoric heating stroke in which the working medium draws an amount of heat, $Q_{\mathrm{in}}$, from the hot reservoir while the external field is held constant. The third stroke ($\mathrm{C}\rightarrow \mathrm{D}$) is an isentropic expansion where the working medium is again disconnected from the thermal reservoirs and the external field is varied from $B_2$ back to $B_1$. During this stroke an amount of work, $W_{\mathrm{exp}}$, is extracted from the expansion of the working medium. The final stroke ($\mathrm{D}\rightarrow \mathrm{A}$) is an isochoric cooling stroke in which the working medium expels and amount of heat, $Q_{\mathrm{out}}$ to the cold reservoir while the external field is held constant. Note that the work parameter ($B$) plays the role of an \textit{inverse} volume, increasing during the compression stroke ($\mathrm{A} \rightarrow \mathrm{B}$) and decreasing during the expansion stroke ($\mathrm{C} \rightarrow \mathrm{D}$). 

Characteristic of the framework of endoreversibility, we will assume the working medium remains in a state of local equilibrium throughout the cycle, but, due to finite-time thermalization strokes, never achieves global equilibrium with the reservoirs. The thermodynamic equation of state for the internal energy of working medium at each corner of the cycle must thus be expressed in terms of the corresponding temperature, $T_A$, $T_B$, $T_C$, or $T_D$, and the external field strength, $B_1$ or $B_2$. Ultimately, we want to determine expressions for the engine performance figures of merit      solely in terms of the experimentally controllable parameters, namely the temperatures of the thermal reservoirs, $T_l$ and $T_h$, the magnetic field strengths $B_1$ and $B_2$, and the durations of the heating and cooling strokes, $\tau_h$ and $\tau_l$. In order to do so, we must model the thermal conduction during the isochoric strokes and apply the constraint that the entropy remains constant during the isentropic strokes. For this endoreversible analysis we will follow the procedure established in Ref. \cite{deffner2018efficiency}.

During the isentropic compression stroke ($\mathrm{A} \rightarrow \mathrm{B}$) the working medium is decoupled from the thermal reservoirs. As such, all change in the working medium's internal energy can be associated with work,
\begin{equation}
	\label{eq:Wcomp}
	W_{\mathrm{comp}} = U_{\mathrm{B}}(T_{\mathrm{B}}, B_2) - U_{\mathrm{A}}(T_{\mathrm{A}}, B_1).
\end{equation} 

During the isochoric heating stroke ($\mathrm{B} \rightarrow \mathrm{C}$), the external field is held constant. Thus the difference in internal energy can be associated entirely with heat,  
\begin{equation}
\label{qinendo}
Q_{\mathrm{in}}=U_{\mathrm{C}}(T_{\mathrm{C}},B_{2})-U_{\mathrm{B}}(T_{\mathrm{B}},B_{2}).
\end{equation}
As mentioned above, unlike in the quasistatic case, $T_{\mathrm{C}} \neq T_h$ since the working medium does not fully thermalize with the hot reservoir. As the heating stroke is now carried out in finite time, we must determine how the temperature of the working medium changes during the duration of the stroke. The temperatures $T_{\mathrm{B}}$ and $T_{\mathrm{C}}$, corresponding to the temperature of the working medium at the beginning and ending of the heating stroke, respectively, must satisfy the conditions,
\begin{equation}
    T(0)= T_{\mathrm{B}}, \quad T(\tau_{h})= T_{\mathrm{C}} \quad \mathrm{and} \quad T_{\mathrm{B}} < T_{\mathrm{C}} \leq T_{h},
\end{equation}
where $\tau_{h}$ is the duration of the heating stroke. Consistent with the assumptions of endoreversibility, we model the temperature change from $T_{\mathrm{B}}$ to $T_{\mathrm{C}}$ using Fourier's law,
\begin{equation}
\label{isochoricheatingdiff}
    \frac{d T}{d t} =-\alpha_{h}\left(T(t)- T_{h}\right),
\end{equation}
where $\alpha_{h}$ is a constant that depends on the thermal conductivity and heat capacity of the working medium. Solving Eq. (\ref{isochoricheatingdiff}) yields, 
\begin{equation}
\label{eq:isoheating}
    T_{\mathrm{C}} - T_{h}= (T_{\mathrm{B}} - T_{h}) e^{-\alpha_{h}\tau_{h}}.
\end{equation}

Just as in the compression stroke, the work extracted during the isentropic expansion stroke ($\mathrm{C} \rightarrow \mathrm{D}$) is found from,
\begin{equation}
	\label{eq:Wexp}
	W_{\mathrm{exp}} = U_{\mathrm{D}}(T_{\mathrm{D}}, B_1) - U_{\mathrm{C}}(T_{\mathrm{C}}, B_2).
\end{equation} 

During the isochoric cooling stroke ($\mathrm{D} \rightarrow \mathrm{A}$) the heat exchanged with the cold reservoir is given by,
\begin{equation}
\label{qoutendo}
Q_{\mathrm{out}}=U_{\mathrm{A}}(T_{\mathrm{A}},B_{1})-U_{\mathrm{D}}(T_{\mathrm{D}},B_{1}),
\end{equation}
where, in analogy to the heating stroke, $T_{\mathrm{A}}$ and $T_{\mathrm{D}}$ satisfy the conditions,
\begin{equation}
    T(0)= T_{\mathrm{D}} \quad \mathrm{and} \quad T(\tau_{l})= T_{\mathrm{A}} \quad \mathrm{with} \quad T_{\mathrm{D}} > T_{\mathrm{A}} \geq T_{l}.
\end{equation}
We again apply Fourier's law to model the temperature change during the stroke, 
\begin{equation}
\label{isochoriccoolingdiff}
    \frac{d T}{d t} =-\alpha_{l}\left(T(t)- T_{l}\right),
\end{equation}
which after solving yields,
\begin{equation}
\label{eq:isocooling}
    T_{\mathrm{A}} - T_{l} =\left(T_{\mathrm{D}} - T_{l}\right)e^{-\alpha_{l}\tau_{l}}.
\end{equation}

With expressions for the work done and heat exchanged during each stroke of the cycle we can now determine the cycle efficiency,
\begin{equation}
	\label{eq:eff}
	\eta = -\frac{W_{\mathrm{comp}}+W_{\mathrm{exp}}}{Q_{\mathrm{in}}},
\end{equation}
and power output,
\begin{equation}
	\label{eq:P}
	P = -\frac{W_{\mathrm{comp}}+W_{\mathrm{exp}}}{\gamma (\tau_h + \tau_l)}.
\end{equation}
Note that $\gamma$ is a multiplicative factor that implicitly incorporates the duration of the isentropic strokes \cite{deffner2018efficiency}.

\begin{figure*}
	\subfigure[]{
		\includegraphics[width=.28\textwidth]{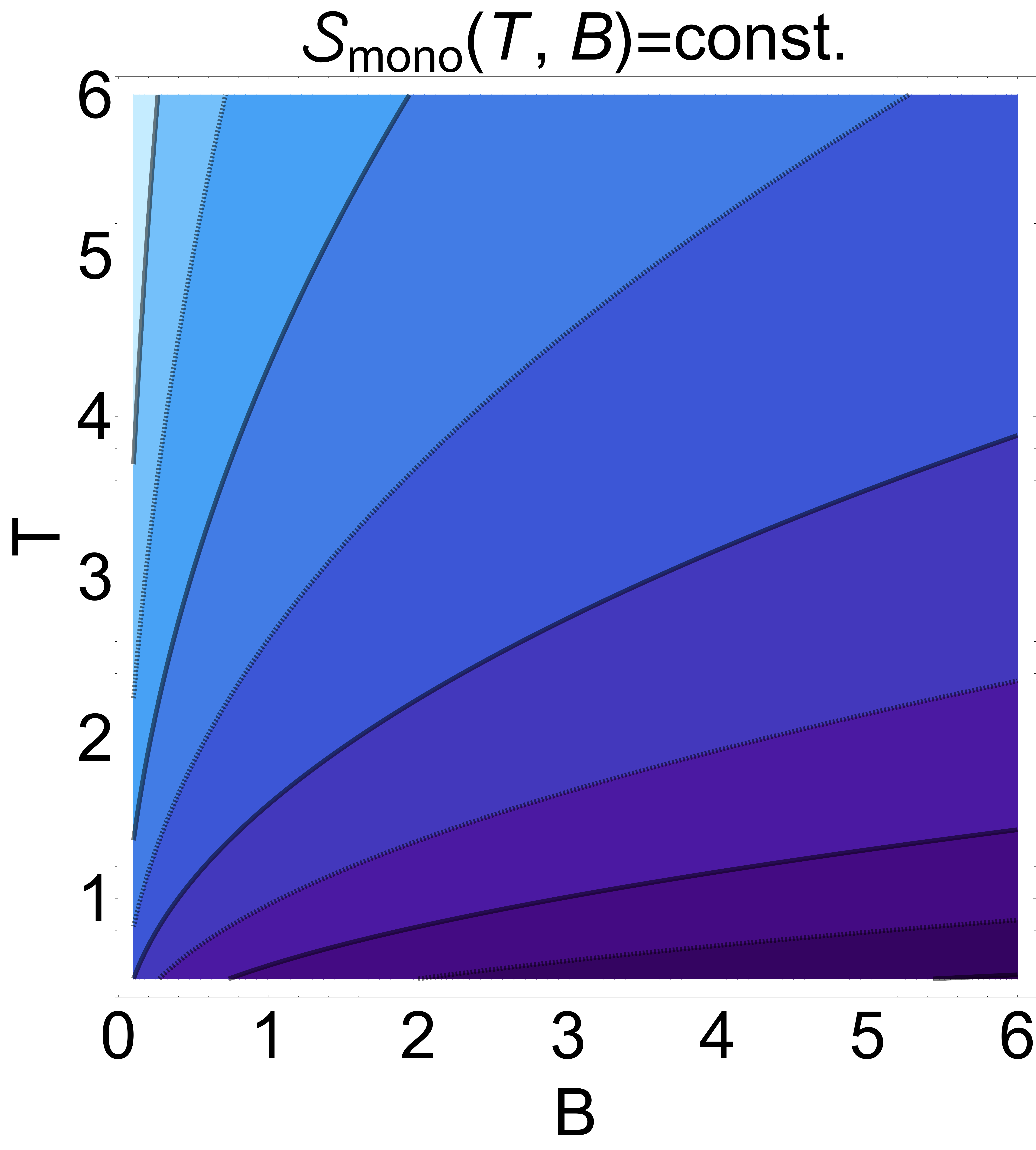}
	}
	\hspace{5mm}
	\subfigure[]{
		\includegraphics[width=.28\textwidth]{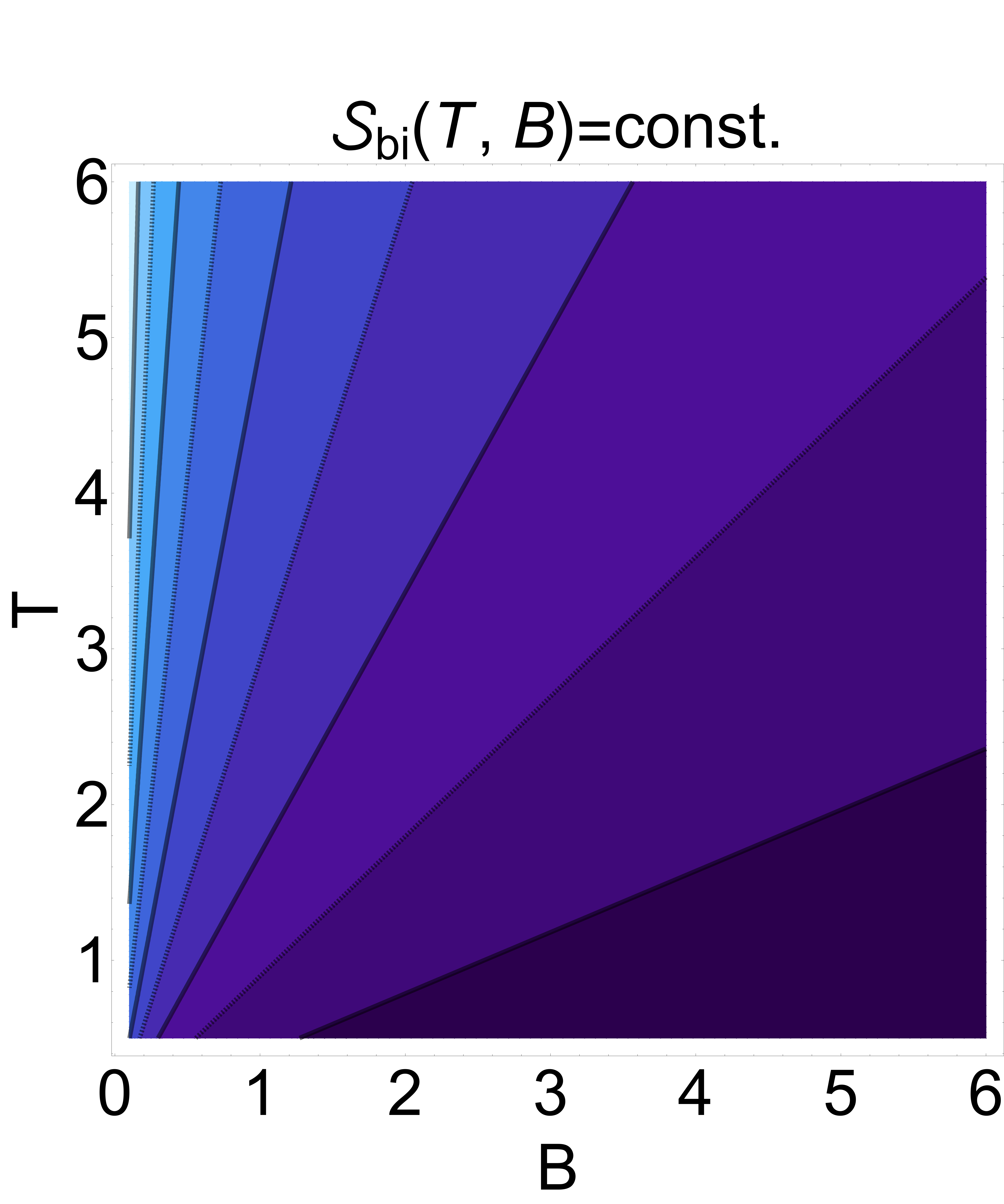}
	}
	\hspace{5mm}
	\subfigure[]{
		\includegraphics[width=.28\textwidth]{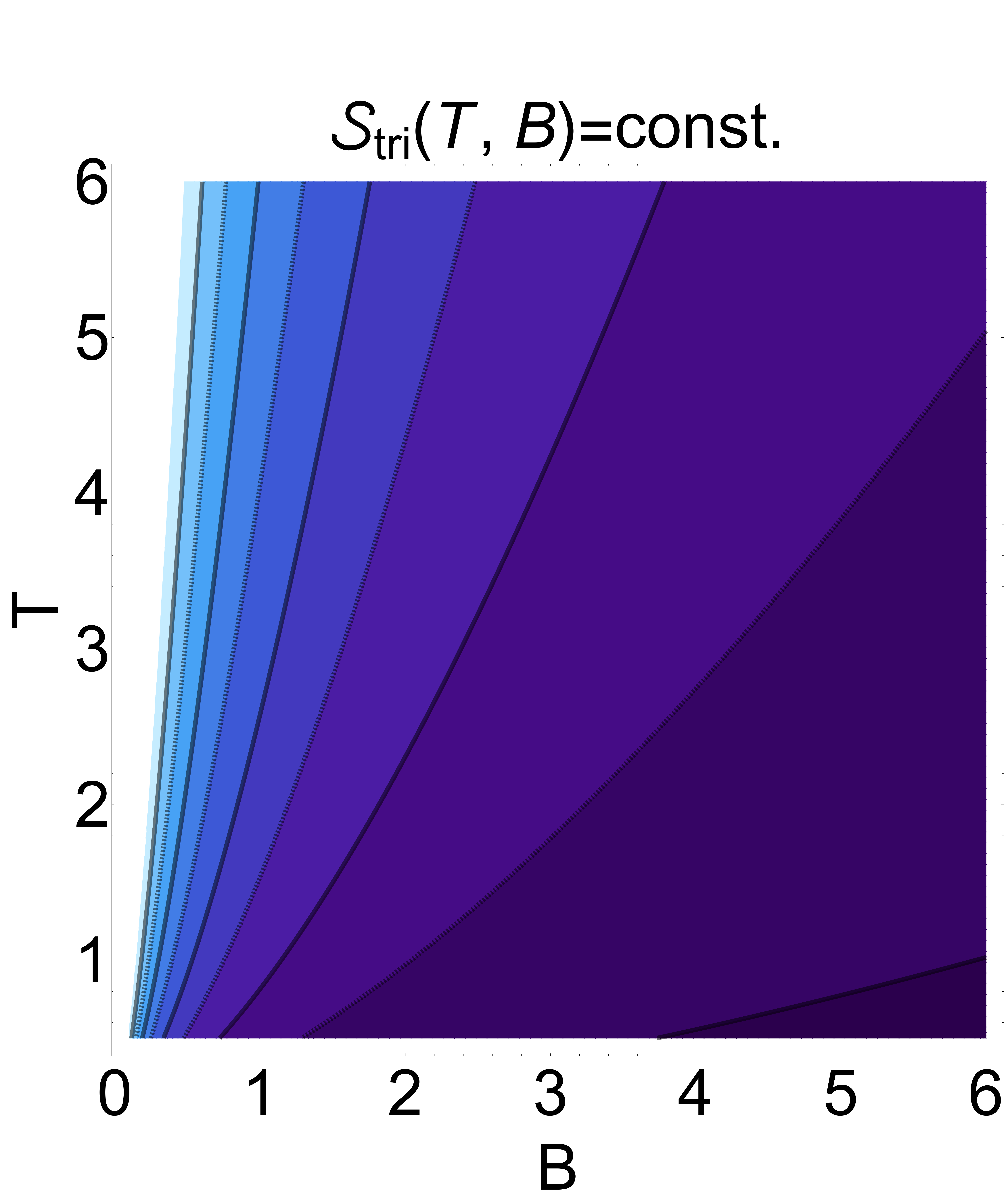}
	}
    \caption{Isentropic curves as a function of the temperature and external field for (a) monolayer, (b) bilayer, and (c) trilayer systems. Darker shading indicates lower entropy. Here we have dimensionless parameters with $k_B = 1$.}
    \label{fig:svsb}
\end{figure*}

By definition, the entropy remains constant during the isentropic strokes. We can use this fact to obtain a relationship between the initial and final temperatures and magnetic field strengths during the isentropic strokes. Using $dS(T,B) = 0$ we obtain the following first order differential equation,
\begin{equation}
\label{differential}
    \frac{dB}{dT}=-\frac{\left(\frac{\partial S}{\partial T}\right)_{B}}{\left(\frac{\partial S}{\partial B}\right)_{T}}.
\end{equation}
Taking the partial derivatives of the entropy found from Eq. \eqref{eq:FandS} we arrive at, 
\begin{equation}
    \label{eq:dBdT}
    \frac{d B}{d T} = \frac{2 B}{\mathcal{N} T}.
\end{equation}
Solving Eq. \eqref{eq:dBdT} for the compression stroke we find,
\begin{equation}
    \label{eq:endorel1}
    \frac{T_{\mathrm{A}}}{T_{\mathrm{B}}} = \left(\frac{B_1}{B_2}\right)^{\frac{\mathcal{N}}{2}}.
\end{equation}
Similarly, solving Eq. \eqref{eq:dBdT} for the expansion stroke we have,
\begin{equation}
    \label{eq:endorel2}
    \frac{T_\mathrm{C}}{T_{\mathrm{D}}} = \left(\frac{B_2}{B_1}\right)^{\frac{\mathcal{N}}{2}}.
\end{equation}
This relationship between the temperature, external field, and number of layers can be seen graphically in Fig. \ref{fig:svsb}, where we have plotted curves of constant entropy as a  function of the temperature and external field for monolayer, bilayer, and trilayer graphene.  

\section{ Results}
\label{sec:5}

\subsection{Efficiency}

We are now in a position where we can determine all of our characterizations of engine performance in terms of experimentally controllable parameters. First, combining Eq. (\ref{eq:eff}) with Eqs. (\ref{qinendo}), (\ref{eq:Wexp}), (\ref{eq:Wcomp}), (\ref{eq:endorel1}) and (\ref{eq:endorel2}) we arrive at a simple expression for the engine efficiency, 
\begin{equation}
    \label{eq:efficiency}
    \eta = 1 - \left(\frac{B_1}{B_2}\right)^{\frac{\mathcal{N}}{2}}.
\end{equation}
We note that this expression is strikingly similar to the classical expression for the Otto efficiency, with the layer number $\mathcal{N}$ playing the role of the ratio of heat capacities. 

\subsection{Power}

Similarly, by combining Eq. (\ref{eq:P}) with (\ref{eq:Wexp}), (\ref{eq:Wcomp}), (\ref{eq:endorel1}) and (\ref{eq:endorel2}) we arrive at an analytical expression for the power output,
\begin{equation}
    \label{eq:FullP}
    \begin{split}
    P = & \frac{2 \left(1-\kappa ^{-\mathcal{N}/2}\right)}{\mathcal{N} \gamma \left(\tau_l+\tau_h\right)} \Bigg[\frac{\Sigma }{B_2 \kappa  (\mathcal{N}-1) \Sigma ^{-2/\mathcal{N}}+\Gamma \left(\frac{\mathcal{N}+2}{\mathcal{N}}\right)} \\ 
    &-\frac{\Lambda }{B_2 \kappa  (\mathcal{N}-1) \Lambda ^{-2/\mathcal{N}}+\Gamma \left(\frac{\mathcal{N}+2}{\mathcal{N}}\right)}\Bigg]\Gamma \left(\frac{\mathcal{N}+2}{\mathcal{N}}\right)
   \end{split}
\end{equation}
where we have defined,
\begin{equation}
    \begin{split}
    & \Sigma \equiv  \frac{e^{\alpha_h \tau_h}\left(e^{\alpha_l \tau_l}-1\right)k_B T_l + \left(e^{\alpha_h \tau_h}-1\right)k_B T_h \kappa^{\mathcal{N}/2}}{e^{\alpha_l \tau_l + \alpha_h \tau_h}-1},\\ &
    \Lambda \equiv \frac{\left(e^{\alpha_l \tau_l}-1\right)k_B T_l + e^{\alpha_l \tau_l}\left(e^{\alpha_h \tau_h}-1\right)k_B T_h \kappa^{\mathcal{N}/2}}{e^{\alpha_l \tau_l + \alpha_h \tau_h}-1}.
    \end{split}
\end{equation}
Examining Eq. (\ref{eq:FullP}), we see that the power will vanish under the condition that $\Sigma = \Lambda$. This occurs under three conditions. The first is that $\kappa^{\mathcal{N}/2} \rightarrow T_l/T_h$. We see from Eq. \ref{eq:efficiency} that this corresponds to limit of Carnot efficiency, under which we would expect the power to vanish. The second and third conditions are when $\exp{(\alpha_h \tau_h)} \rightarrow 1$ and $\exp{(\alpha_l \tau_l)} \rightarrow 1$, respectively. These conditions correspond to the limits of instantaneous thermalization strokes or vanishing thermal conductivity, both of which would prevent heat transfer and thus result in zero power. We also note that the power vanishes in the quasistatic limit of $\tau_l + \tau_h \rightarrow \infty$. From Eqs. (\ref{eq:isoheating}) and (\ref{eq:isocooling}) we see that this limit yields $T_3 = T_h$ and $T_1 = T_l$, which in turn maximizes the efficiency. This is a demonstration of the well-established trade-off between efficiency and power.            

\begin{figure*}
	\subfigure[]{
		\includegraphics[width=.3\textwidth]{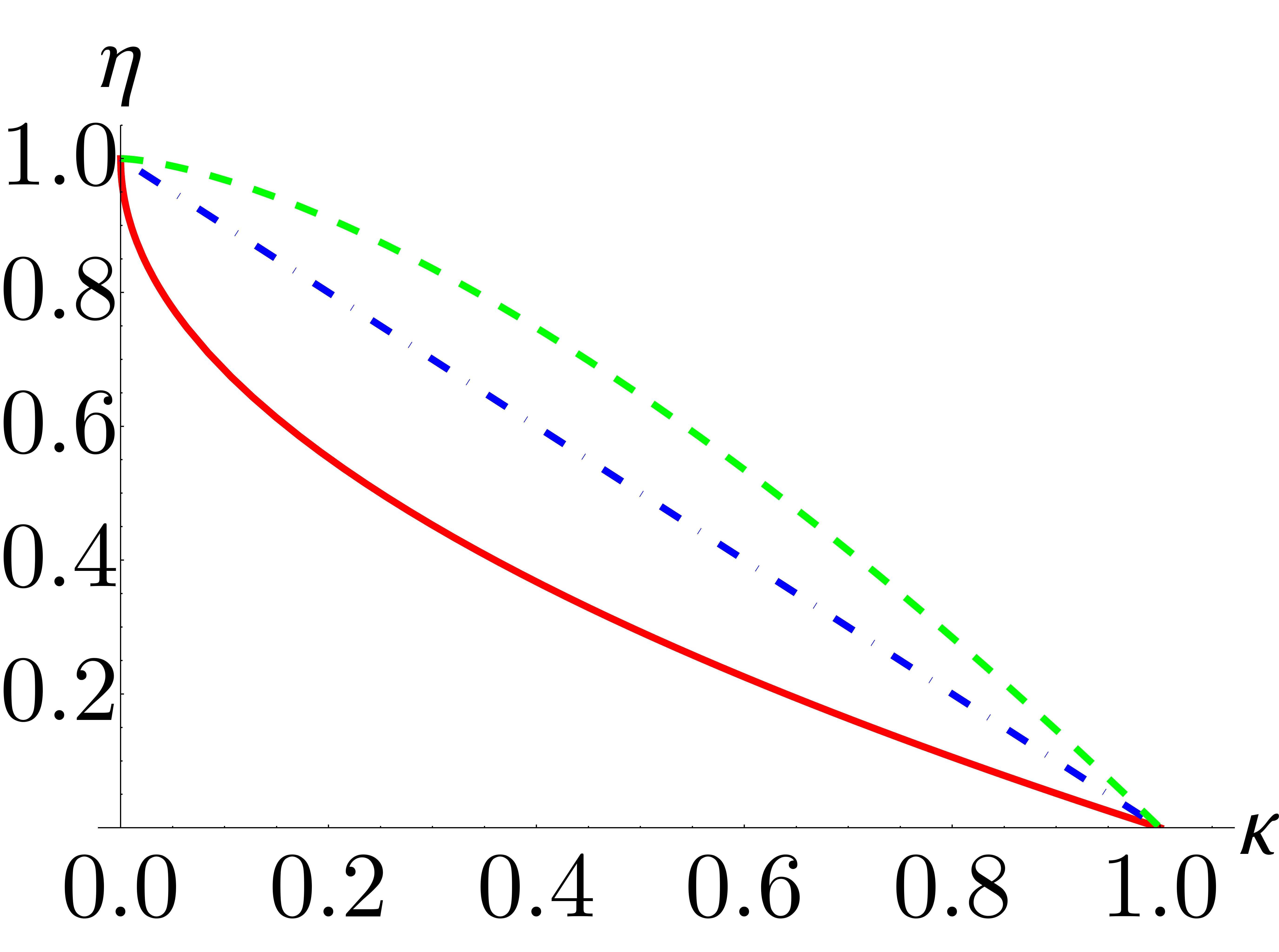}
	}
	\hspace{5mm}
	\subfigure[]{
		\includegraphics[width=.3\textwidth]{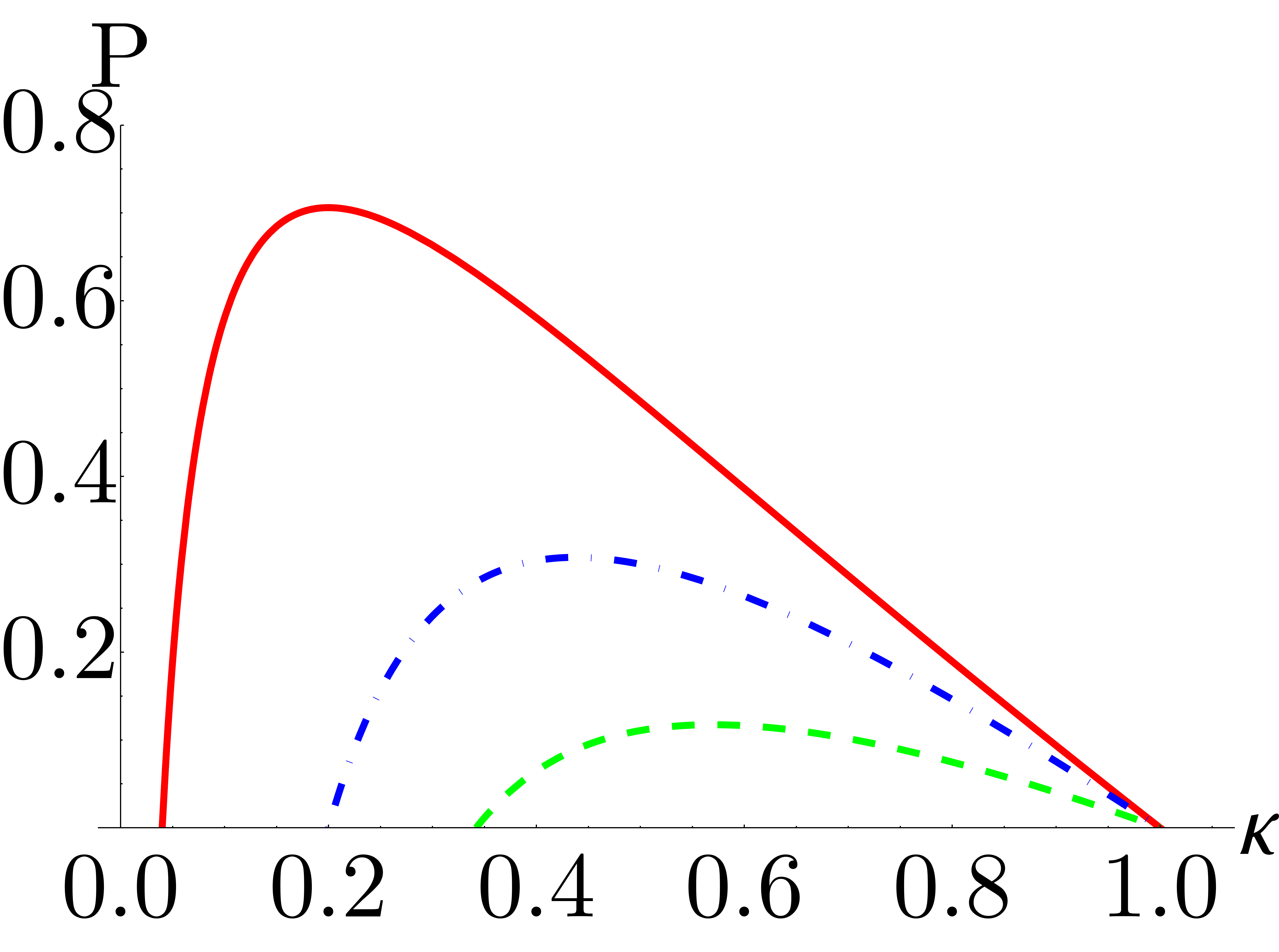}
	}
    \caption{(a) Efficiency and (b) power as a function of the compression ratio for monolayer (red, solid), bilayer (blue dot-dashed), and trilayer (green, dashed) working mediums. Parameters for figure (b) are $B_2 = 2$, $T_h = 5$, $T_l = 1$, and $\alpha_l = \alpha_h = \tau_l = \tau_h =1$.}
    \label{fig:effPplots}
\end{figure*}

The efficiency and power are plotted as a function of the compression ratio, $\kappa$, in Fig. \ref{fig:effPplots} for the monolayer, bilayer, and trilayer systems. We see that the monolayer system has the lowest efficiency, but highest power output while the opposite is true for the trilayer system. The efficiency and power of the bilayer system falls between the monolayer and trilayer results.  

\subsection{Efficiency at maximum power}

Due to the inherent trade-off between efficiency and power mentioned above, efficiency alone does not provide the most practically useful metric of engine performance. Instead, this role is played by the efficiency at maximum power. In this case the EMP is found by maximizing the power output with respect to the compression ratio, $\kappa$, and then determining the efficiency corresponding to this value of $\kappa$. For a classical Otto cycle, the EMP is given by the Curzon-Ahlborn efficiency,
\begin{equation}
    \label{eq:CA}
    \eta_{\mathrm{CA}} = 1 - \sqrt{\frac{T_l}{T_h}}
\end{equation}

\begin{figure}  
\includegraphics[width=0.45\textwidth]{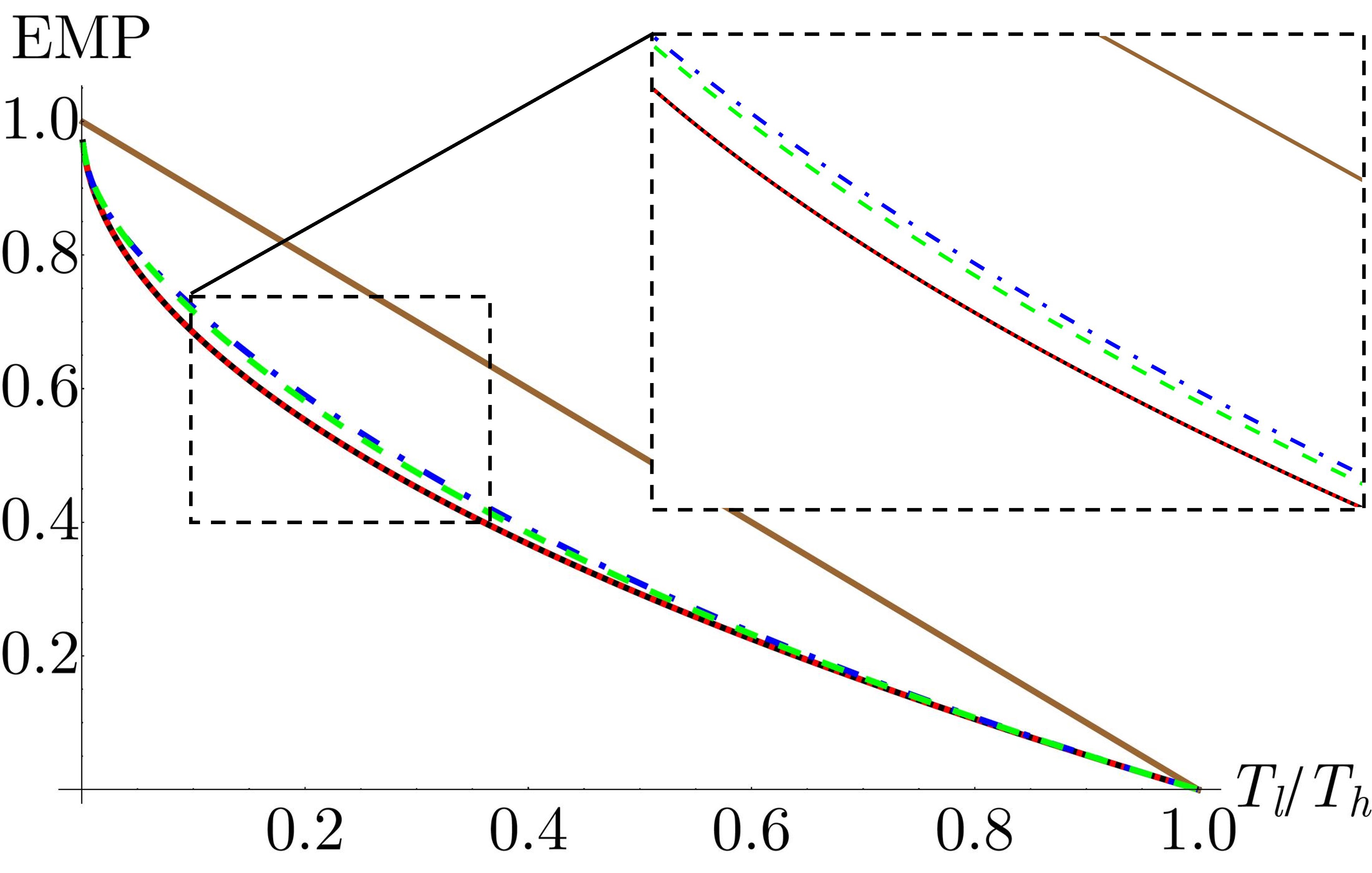}
 \caption{Efficiency at maximum power as a function of the ratio of bath temperatures for monolayer (red, dotted), bilayer (blue, dot-dashed) and trilayer (green, dashed) working mediums. The Carnot (brown, upper solid) and Curzon-Ahlborn (black, lower solid) efficiencies are given for comparison. Parameters are chosen such that the engine is operating in the quantum regime with $\theta_{\mathcal{N}} B_2/k_B T_l = 20$. Other parameters are $\alpha_l = \alpha_h = \tau_l = \tau_h =1$.}
 \label{fig:EMP1}
\end{figure}

Due to the complicated expression for power in Eq. (\ref{eq:FullP}) we maximize the power numerically. The EMP as a function of the ratio of bath temperatures is shown in Fig. \ref{fig:EMP1}. We see that for the monolayer case, the EMP is identical to the Curzon-Ahlborn efficiency. This result can be confirmed analytically by taking the derivative of Eq. (\ref{eq:FullP}) and confirming that it vanishes for $\mathcal{N}=1$ and $\kappa = T_l/T_h$. 

For the bilayer and trilayer systems, however, we see that the EMP exceeds the CA efficiency. The EMP is largest for the bilayer system, decreasing slightly in the trilayer case. This trend continues, with the EMP of larger layer numbers converging back towards the CA efficiency. However, it is important to note that if we increase the number of graphene layers significantly beyond the trilayer case, the assumptions made in determining the closed form of the partition function begin to break down.

It has been previously shown for both classical and quantum working mediums that, within the regime of linear response, EMP is bounded by the CA efficiency \cite{Broeck2005, Esposito2009, Izumida2009, Benenti2017, Singh2022, Chen2022}. To achieve higher EMP requires going beyond the linear regime or by breaking time-reversal symmetry \cite{Allahverdyan2008, Izumida2008, Schmiedl2008, Esposito2009EPL, Sothmann2012, Apertet2012, Jordan2013, Brandner2013, Brandner2013NJP}. For a cyclic engine, the regime of linear response occurs near the equilibrium limit $T_l \approx T_h$. 

\begin{figure*}
    \subfigure[]{
		\includegraphics[width=.29\textwidth]{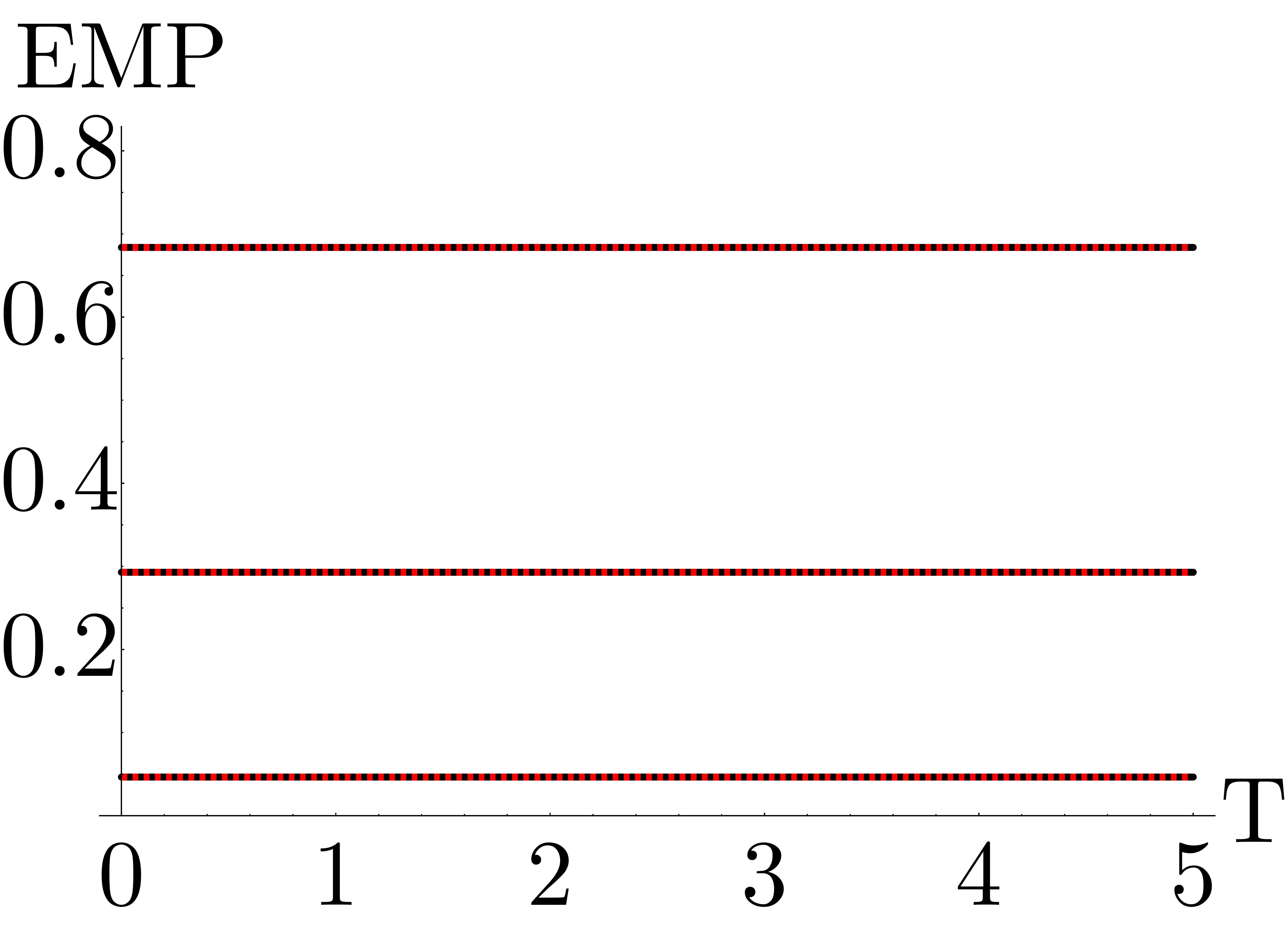}
	}
	\hspace{5mm}
	\subfigure[]{
		\includegraphics[width=.29\textwidth]{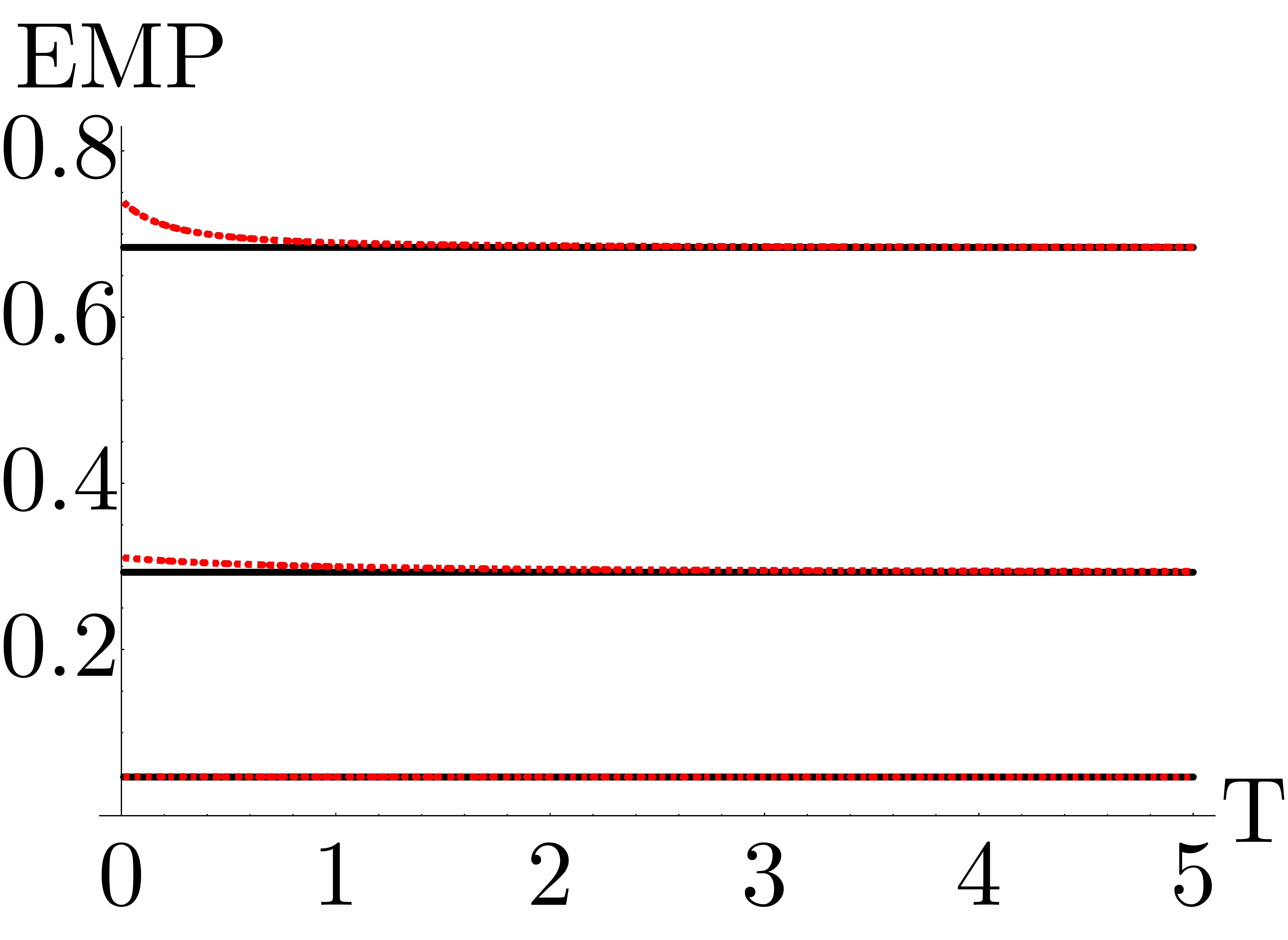}
	}
	\hspace{5mm}
	\subfigure[]{
		\includegraphics[width=.29\textwidth]{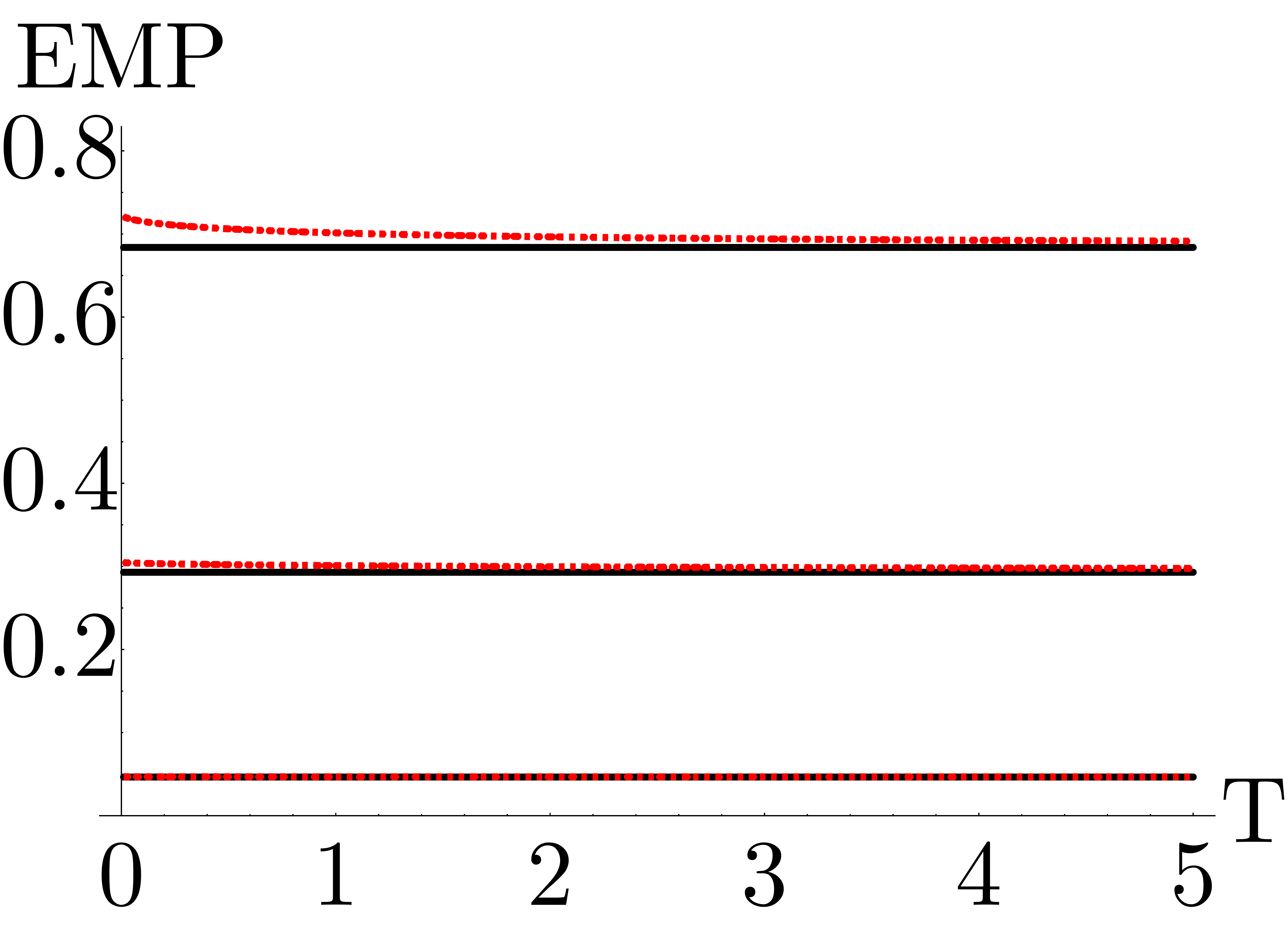}
	}
    \caption{EMP (red, dashed) in comparison to the Curzon-Ahlborn efficiency (black, solid) as a function of temperature for (a) monolayer, (b) bilayer, and (c) trilayer working mediums. The bottom pair of lines in each plot corresponds to $\epsilon = 1.1$, the middle pair to $\epsilon = 2$ and the top pair to $\epsilon = 10$. Here $T_l \equiv T$, $T_h \equiv \epsilon T_l$. Parameters are $\alpha_l = \alpha_h = \tau_l = \tau_h =1$ and $B_2 = 2$.}
    \label{fig:LREMPplots}
\end{figure*}

To probe the behavior of the EMP for a multilayer graphene working medium in and around the linear response regime we define $T_l \equiv T$ and $T_h \equiv \epsilon T$.  In Fig. \ref{fig:LREMPplots} we plot the EMP for the monolayer, bilayer and trilayer working mediums in comparison to the CA efficiency for $\epsilon = 1.1$, $2$, and $10$. As expected, for the monolayer system we see that at all examined values of $\epsilon$ the EMP is identical to the CA efficiency. For the bilayer and trilayer systems we see that at $\epsilon = 1.1$, close to the equilibrium limit, the EMP is identical to CA, consistent with the results in the works mentioned above. As we move away from the equilibrium limit by increasing $\epsilon$ we see that, at low bath temperatures, the EMP exceeds CA, but that as the temperature increases the EMP converges back to the CA efficiency. As observed in Fig. \ref{fig:EMP1}, in the low temperature regime the bilayer EMP exceeds the CA efficiency by a greater amount than the trilayer working medium. However, as temperature increases the bilayer EMP converges to CA faster than the trilayer EMP. 

From these results we see that two conditions must be met for the EMP to exceed CA. First, the difference in bath temperatures must be sufficiently far from the equilibrium limit of $T_l \approx T_h$. Second, the cycle must be operating in the low-temperature, quantum regime which for the multilayer graphene working medium is determined by the condition $\theta_{\mathcal{N}} B_2/k_B T_l \gg 1$. This second condition is consistent with the results shown in Ref. \cite{deffner2018efficiency} for harmonic working mediums.    

\subsection{Engine vs refrigerator}

For any arbitrary choice of parameters it is not guaranteed that the Otto cycle will function as an engine. In general, there are four possible types of thermal machines, corresponding to all possible combinations of directions heat and work flow consistent with the first and second laws of thermodynamics. An engine corresponds to positive work output, along with heat flow from the hot bath into the working medium, and from the working medium into the cold bath. A refrigerator corresponds to negative work output, along with heat flow from the cold bath into the working medium and from the working medium into the hot bath. A heater corresponds to negative work output, along with heat flow from the working medium into both baths. Finally, an accelerator corresponds to negative work output along with heat flow from the hot bath into the working medium and from the working medium into the cold bath. 

By examining the signs of Eqs. (\ref{qinendo}), (\ref{eq:Wexp}), (\ref{qoutendo}), and (\ref{eq:Wcomp}) across the parameter space we can determine the regions where cycle will function as each type of thermal machine. Note, both the heater and accelerator are fundamentally nonequilibrium devices and thus we would not expect to find regions of parameter space corresponding to these devices under the assumption of endoreversible behavior. 

\begin{figure*}
	\subfigure[]{
		\includegraphics[width=.25\textwidth]{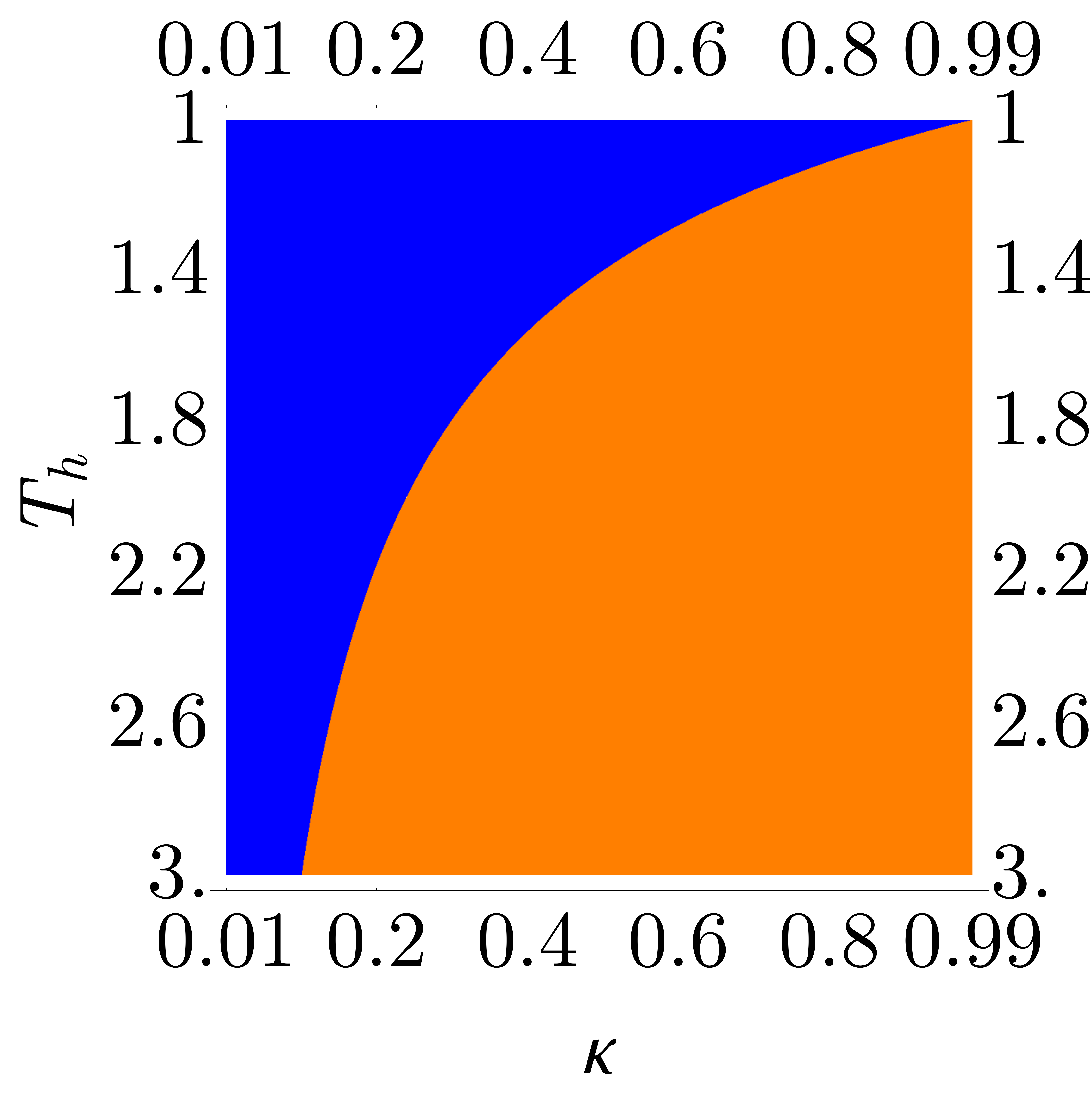}
	}
	\hspace{5mm}
	\subfigure[]{
		\includegraphics[width=.25\textwidth]{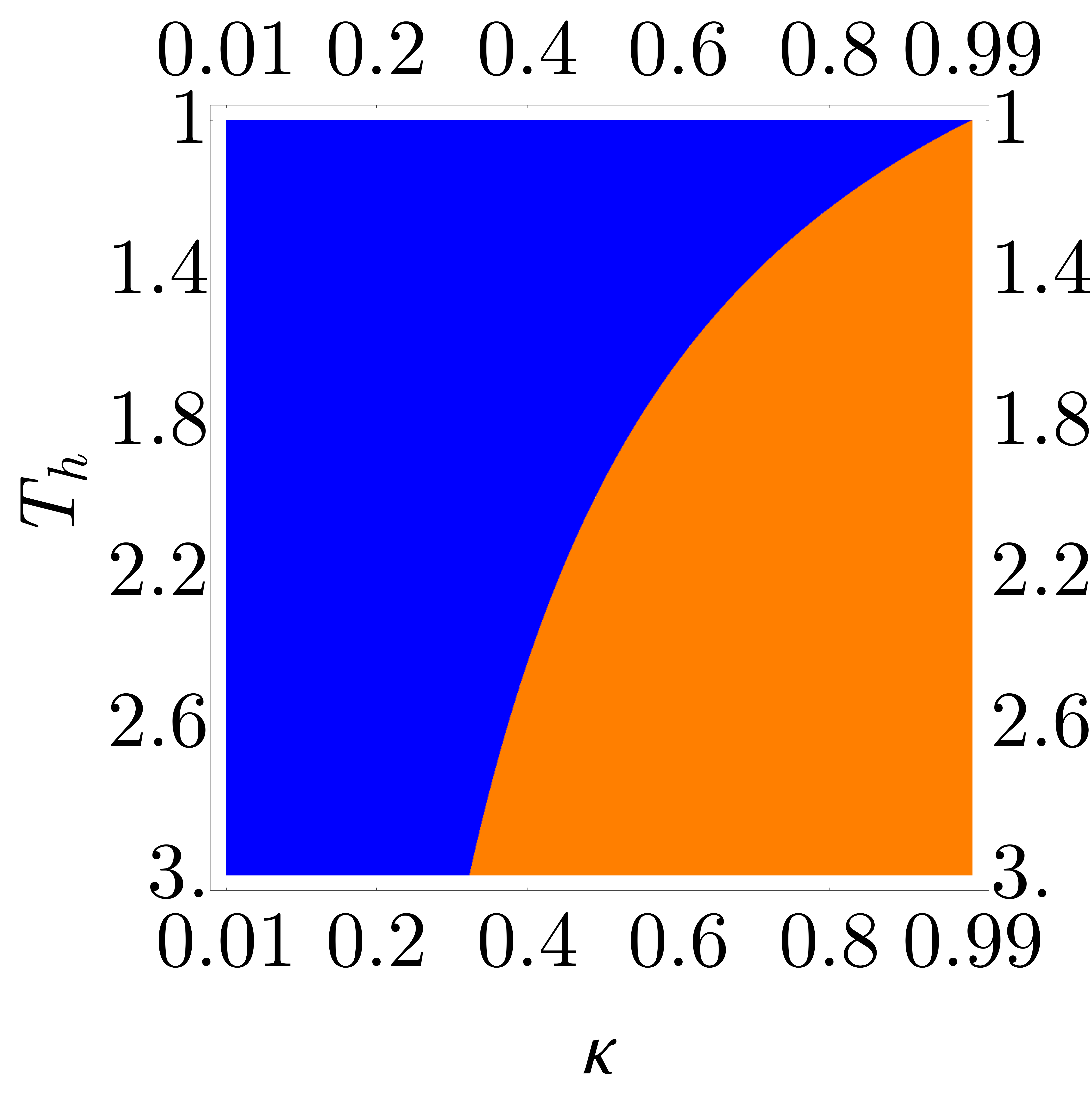}
	}
	\hspace{5mm}
	\subfigure[]{
		\includegraphics[width=.25\textwidth]{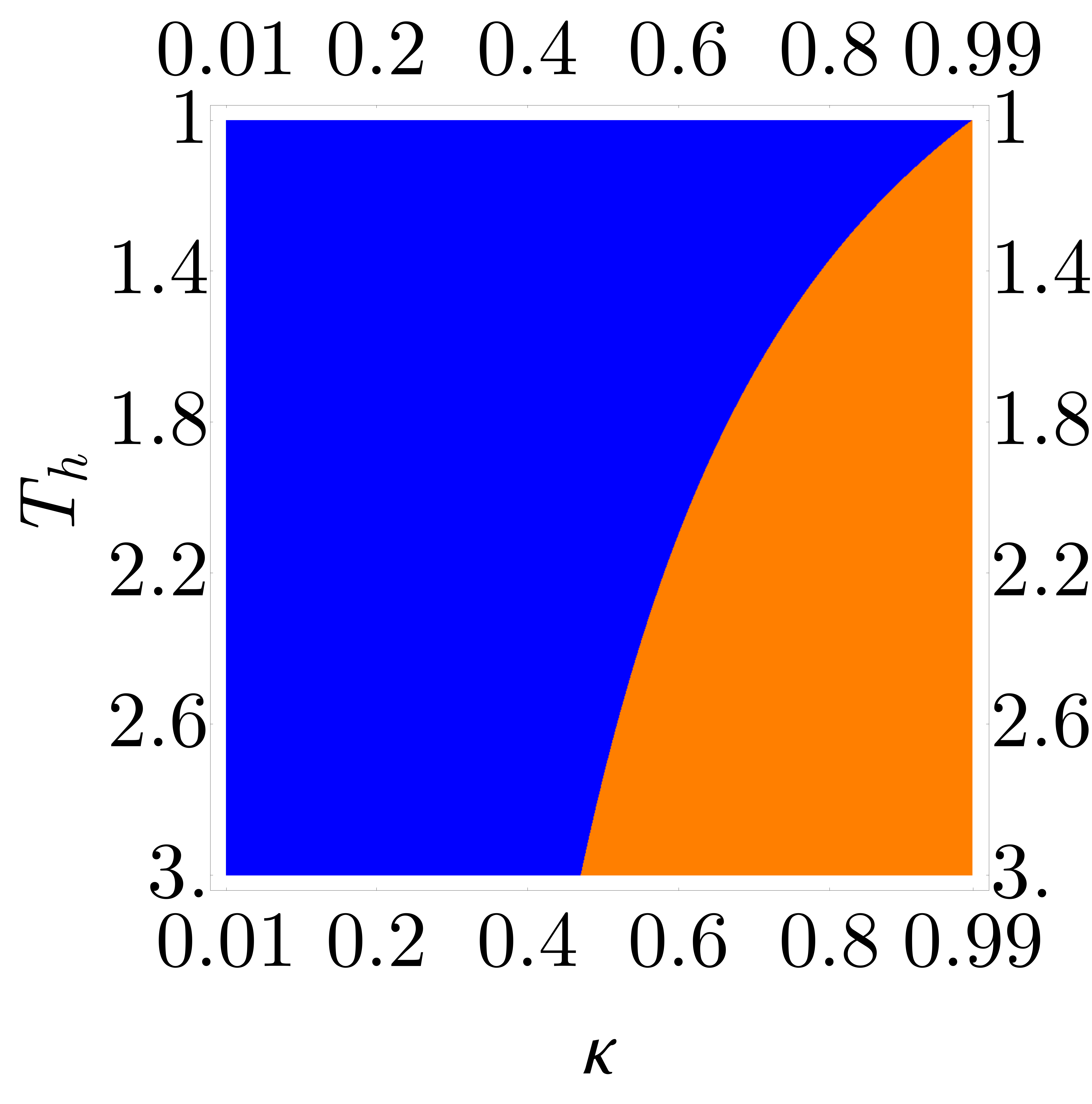}
	}
    \caption{Regions of parameter space where the endoreversible cycle functions as an engine (orange, convex) and refrigerator (blue, concave) for (a) monolayer, (b) bilayer, and (c) trilayer working medium. Parameters are $B_2 = 10$, $\alpha_l = \alpha_h = \tau_l = \tau_h =1$, and $T_l =1$.}
    \label{fig:regionsengine}
\end{figure*}

In Fig. \ref{fig:regionsengine} we show the regions where the cycle functions as either an engine or a refrigerator as a function of the hot bath temperature and compression ratio. We see that over the same region of parameter space the layer number has a significant impact on the apportionment between engine and refrigerator. In the monolayer case we see that majority of the examined region corresponds to the engine regime, while in the trilayer case the opposite is true, with a larger portion of the explored space corresponding to the refrigerator regime. The origins of this behavior can be understood from the plot of the power in Fig. \ref{fig:effPplots}. As the layer number decreases, we see that the reduced power output leads to a reduced region of positive work, and thus a smaller engine regime. 

\section{Conclusions}
\label{sec:6}

We have examined the performance of an endoreversible Otto cycle with a working medium of a multilayer graphene system. We have found that all examined performance metrics, including the efficiency, power, EMP and parameter regions under which the cycle functions as an engine or refrigerator all depend significantly on the number of layers. Most notably we have found that the EMP for bilayer and trilayer graphene working mediums exceeds the Curzon-Ahlborn efficiency. Conversely, we have demonstrated that the EMP of a monolayer graphene working medium is identical to the CA efficiency. As the energy spectrum of monolayer graphene can be mapped to that of the relativistic Dirac oscillator, this result is consistent with previous work examining the performance of endoreversible Otto engines with relativistic oscillators as the working medium \cite{Myers2021NJP}. We have also found two conditions necessary for the graphene EMP to exceed the CA efficiency, namely the cycle must be sufficiently far from the equilibrium limit and must be operating in the low temperature regime corresponding to $\theta_{\mathcal{N}} B_2/k_B T_l \gg 1$. 

Experimental implementation of the multilayer graphene engine requires precise control of the number of layers as well as a tunable magnetic field of sufficient strength to induce Landau quantization. Fine control over multilayer structures has been demonstrated by folding monolayer graphene nanoribbons \cite{Chen2019} and precise multilayer thickness measurements can be accomplished with electron spectroscopy \cite{Xu2010}. Strong external magnetic fields can be generated either by direct application \cite{Zhang2006} or through strain induced psuedo-magnetic fields \cite{Guinea2010, Levy2010}. 

\begin{acknowledgments}
The authors would like to thank Oscar Negrete for enlightening discussions in the early stages of this work. N.M.M. acknowledges support from AFOSR (FA2386-21-1-4081, FA9550-19-1-0272, FA9550-23-1-0034) and ARO (W911NF2210247, W911NF2010013). F.J.P. acknowledges financial support from ANID Fondecyt, Iniciación en Investigación 2020 grant No. 11200032,  ANID Fondecyt grant No. 1210312, ``Millennium Nucleus in NanoBioPhysics” project NNBP NCN2021\textunderscore 021 and USM-DGIIE. N.C. acknowledges support from ANID Fondecyt Iniciación en Investigación Project No. 11221088 and IAI-UTA, and the hospitality of Ohio University. P.V. acknowledges support from ANID Fondecyt grant No. 1210312 and ANID PIA/Basal grant No. AFB180001.
\end{acknowledgments}

\bibliography{grapheneEngine}

\end{document}